\documentclass[slac_one]{revtex4}
\usepackage{graphicx}
\usepackage{fancyhdr}
\begin{document}

\title{{\small{Hadron Collider Physics Symposium (HCP2008),
Galena, Illinois, USA}}\\ 
\vspace{12pt}
Results and prospects for {\boldmath $\Upsilon$}(5S) running at {\boldmath $B$}-factories} 

%

\author{A. Drutskoy}
\affiliation{University of Cincinnati, Cincinnati, OH 45221, USA}

\begin{abstract}
Recent results and future prospects for $\Upsilon$(5S) running at 
$B$-factories are discussed. The first Belle measurements with
23.6\,fb$^{-1}$ of data taken at the $\Upsilon$(5S) energy are
reported. Eligibility of potential measurements expected with 
100\,fb$^{-1}$ and 1000\,fb$^{-1}$ of data at the $\Upsilon$(5S) 
is estimated.
\end{abstract}

\maketitle

\thispagestyle{fancy}


\section{INTRODUCTION} 
During the last several years an opportunity for $B_s^0$ meson studies at the 
$e^+ e^-$ colliders running at the $\Upsilon$(5S) resonance 
has been extensively explored. The first evidence for $B_s^0$ production at the
$\Upsilon$(5S) was found by the CLEO collaboration \cite{cleoi,cleoe} using
a dataset of 0.42\,fb$^{-1}$ collected in 2003. This study indicated
that practical $B_s^0$ measurements at the $\Upsilon$(5S) are possible
with a dataset of at least 20\,fb$^{-1}$, which can be easily collected at
$B$ factories running with $\sim$10$^{34}$cm$^{-2}$sec$^{-1}$ luminosity. 
To test the feasibility of a $B_s^0$ physics program the Belle collaboration 
collected at the $\Upsilon$(5S) a dataset of 1.86\,fb$^{-1}$ in 2005. 
After the successful analysis of these data \cite{beli,bele}, Belle collected 
a bigger sample of 21.7\,fb$^{-1}$ in 2006.

\section{EVENT CLASSIFICATION AND FULL {\boldmath $B_S^0$} EVENT NUMBER DETERMINATION AT THE {\boldmath $\Upsilon$(5S)} RESONANCE}

Several $b\bar{b}$ resonances have been observed in the $e^+ e^-$
hadronic cross-section in the energy region $\sim$10 GeV \cite{pdg}.
Among these resonances, the $\Upsilon$(4S) has a mass 
slightly above the $B\bar{B}$ production threshold and
decays to $B^+B^-$ or $B^0\bar{B}^0$ pairs with almost 100$\%$ probability.
The next $\Upsilon$(5S) resonance has a mass exceeding the
$B_s^0\bar{B}_s^0$ production threshold and can potentially decay 
to various final states with the $B^+$, $B^0$ or $B_s^0$ mesons.
If the $B_s^0$ production rate at the $\Upsilon$(5S) is not small, 
the $\Upsilon$(5S) could play a similar role for comprehensive $B_s^0$ studies 
that the $\Upsilon$(4S) has played for $B^+$ and $B^0$ studies.

Hadronic events produced at the energy region of the $\Upsilon$(5S)
can be classified into the three categories (Fig.~1):
$\Upsilon$(5S) resonance events, $b\bar{b}$ continuum events, 
and $u\bar{u}, d\bar{d}, s\bar{s}, c\bar{c}$ continuum events.
The $\Upsilon$(5S) resonance events and the $b\bar{b}$ continuum 
events (contributions from these two sources are expected to be about
the same) always produce the final states with a $B$ or $B_s$ meson pair,
and, therefore, cannot be topologically separated.
We define the $b\bar{b}$ continuum and $\Upsilon$(5S) events 
collectively as the $b\bar{b}$ events.
All $b\bar{b}$ events are expected to hadronize
in final states with the $B_{(s)}^{(*)}$ meson pair:
$B\bar{B}$, $B\bar{B}^\ast$, $B^\ast\bar{B}$, $B^\ast\bar{B}^\ast$, 
$B\bar{B}\,\pi$, $B\bar{B}^\ast\,\pi$, $B^\ast\bar{B}\,\pi$, 
$B^\ast\bar{B}^\ast\,\pi$, $B\bar{B}\,\pi \pi$,
$B_s^0\bar{B}_s^0$, $B_s^0\bar{B}_s^\ast$, $B_s^\ast\bar{B}_s^0$ or 
$B_s^\ast\bar{B}_s^\ast$. Here $B$ denotes a $B^0$ or a $B^+$ meson and 
$\bar{B}$ denotes a $\bar{B}^0$ or a $B^-$ meson.
The excited states decay to their ground states via $B^* \to B \gamma$ and
$B_s^* \to B_s^0 \gamma$.
All possible final states can be separated into $B_s$
and $B$ channel events; the $B_s$ channel events include
three channels: with zero, one or two $B_s^\ast$ mesons. 

\begin{figure*}[t]
\centering
\includegraphics[width=87mm]{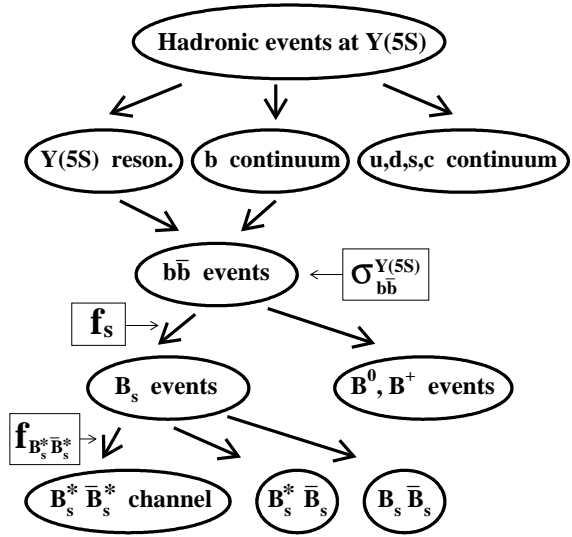}
\vspace{-1.3cm}
\caption{Hadronic event classification at the $\Upsilon$(5S).} \label{JACpic2-f1}
\end{figure*}

In order to measure any $B_s^0$ branching fraction,
the number of $B_s^0$ mesons in a collected $\Upsilon$(5S) data sample
has to be precisely determined. To calculate the $B_s^0$ number
in a dataset with known integrated luminosity $L_{\rm int}$,
the two parameters should be measured in advance:
the total $b\bar{b}$ production cross section at the $e^+ e^-$
center-of-mass energy $\sigma_{b\bar{b}}^{\Upsilon{\rm (5S)}}$,
and the fraction $f_s$ of $B_s^0$ events among all $b\bar{b}$ events.
Then the number of $B_s^0$ mesons in a dataset can be calculated as:
$N_{B_s^0} = 2 \times L_{\rm int} \times 
\sigma_{b\bar{b}}^{\Upsilon{\rm (5S)}} \times f_s$.
For some $B_s^0$ decays with a high background 
level it is reasonable to select events 
only from the $B_s^\ast\bar{B}_s^\ast$ channel. In this case 
the fraction $f_{B_s^\ast\bar{B}_s^\ast}$ of $B_s^\ast\bar{B}_s^\ast$ 
events over all $B_s^{(\ast)}\bar{B}_s^{(\ast)}$ events
should be also measured.

The total $b\bar{b}$ cross 
section $\sigma_{b\bar{b}}^{\Upsilon{\rm (5S)}}$ can be determined
by comparing the number of hadronic events produced at the $\Upsilon$(5S) and 
on the continuum below the $\Upsilon$(4S) energy.
After corrections for the integrated luminosity ratio, the center-of-mass 
energy ratio and the reconstruction efficiency ratio, 
the number of events in continuum dataset will reproduce
the number of $u\bar{u}, d\bar{d}, s\bar{s}$ and $c\bar{c}$ 
continuum events in the $\Upsilon$(5S) dataset.
The dominant systematic uncertainty in the continuum subtraction method
comes from the $\Upsilon$(5S) and continuum luminosity ratio, which
is usually calculated using Bhabha events. In the Belle measurement
\cite{beli} a $\sim 0.4\%$ systematic uncertainty on
the integrated luminosity ratio ${\cal L}_{\rm 5S} / {\cal L}_{\rm cont}$
resulted in the $\sim 5\%$ systematic uncertainty on
$\sigma_{b\bar{b}}^{\Upsilon{\rm (5S)}}$.
Probably with higher statistics, the $\Upsilon$(5S) and continuum dataset 
integrated luminosity ratio can be better measured comparing $D^0$ 
and $D_s$ production at the high momentum region. 
Unfortunately the $\sigma_{b\bar{b}}^{\Upsilon{\rm (5S)}}$ uncertainty 
cannot be essentially reduced below $3\%$ using continuum subtraction 
methods, and it would become a ``core'' uncertainty
on the full $B_s^0$ number determination with increasing statistics.
By now the $\sigma_{b\bar{b}}^{\Upsilon{\rm (5S)}}$ value was measured 
by CLEO \cite{cleoi} to be $(0.301 \pm 0.002 \pm 0.039)$\,nb
and by Belle \cite{beli} to be $(0.302 \pm 0.015)$\,nb.

The $f_s$ value can be determined by comparing $D_s$, $D^0$ or $\phi$
production in $\Upsilon$(5S), $\Upsilon$(4S) and continuum datasets.
This method is based on the fact that the $D_s$ and $\phi$ meson production 
rate is significantly higher in $B_s^0$ decays than in $B^{0/+}$ decays
(and lower for the $D^0$ meson). Unfortunately an additional model 
dependent assumption on the inclusive branching
fraction of $D_s$, $\phi$ or $D^0$ production in $B_s^0$ decays 
has to be made.
Currently the mean PDG value \cite{pdg}, based on CLEO and Belle measurements, 
is $f_s= (19.5^{+3.0}_{-2.3})\%$.
Although the uncertainty can be decreased by about a factor of two with larger
statistics, this method includes a model dependence and has an unavoidable
basic uncertainty. Several model independent methods have been discussed
(using double $D_s$ production, lepton correlations, vertex information,
low momentum photons and so on), however these methods
mostly require rather large statistics or include 
a model dependent parameter.
Generally, we expect to reduce the $f_s$ uncertainty to $\sim(6-8)\%$ with 
currently available statistics of 23.6\,fb$^{-1}$.
The last unknown $f_{B_s^\ast\bar{B}_s^\ast}$ value was recently
measured with improved accuracy: $f_{B_s^\ast\bar{B}_s^\ast}=
(90.3^{+3.8}_{-4.0})\%$ \cite{bdspi}.

The total uncertainty in the number of $B_s^0$ is
expected to come down from $\sim (15-17)\%$ to $\sim (8-9)\%$
due to the Belle statistics increase from 1.86 fb$^{-1}$ to
23.6 fb$^{-1}$. At the moment the most significant uncertainty 
on the $B_s^0$ number comes from the $f_s$ uncertainty. 
It is important to develop 
a robust method to determine $f_s$ in a model independent way.
The uncertainty on the $f_{B_s^\ast\bar{B}_s^\ast}$ value is
smaller than other uncertainties on the number of $B_s^0$ 
and will be further reduced with increasing statistics.

$B_s^0$ signals can be observed using two variables:
the energy difference $\Delta E\,=\,E^{CM}_{B_s^0}-E^{\rm CM}_{\rm beam}$
and the beam-energy-constrained mass
$M_{\rm bc} = \sqrt{(E^{\rm CM}_{\rm beam})^2\,-\,(p^{\rm CM}_{B_s^0})^2}$,
where $E^{\rm CM}_{B_s^0}$ and $p^{\rm CM}_{B_s^0}$ are the energy and momentum
of the $B_s^0$ candidate in the $e^+ e^-$ center-of-mass (CM) system,
and $E^{\rm CM}_{\rm beam}$ is the CM beam energy.
The $B_s^* \bar{B}_s^*$, $B_s^* \bar{B}_s^0$,
$B_s^0 \bar{B}_s^*$ and $B_s^0 \bar{B}_s^0$ intermediate channels 
can be distinguished kinematically in the $M_{\rm bc}$ and $\Delta E$ plane,
where three well-separated $B_s^0$ signal regions
can be defined corresponding to the cases where both, only one, or neither
of the $B_s^0$ mesons originate from a $B_s^*$ decay.
The events obtained from MC simulation of the 
$B_s^0 \to D_s^- \pi^+$ decay are shown in Fig.~2 for the
intermediate $\Upsilon$(5S) decay channels 
$B_s^* \bar{B}_s^*$, $B_s^* \bar{B}_s^0$,
$B_s^0 \bar{B}_s^*$ and $B_s^0 \bar{B}_s^0$.
The signal regions are defined as ellipses
corresponding to $\pm$(2.0--2.5)$\sigma$ (i.e. (95-98)\% acceptance) 
resolution intervals in $M_{\rm bc}$ and $\Delta E$.
The signal events from the different intermediate channels are
well separated in the $M_{\rm bc}$ and $\Delta E$ plane.

\begin{figure*}[h]
\centering
\includegraphics[width=55mm,height=50mm]{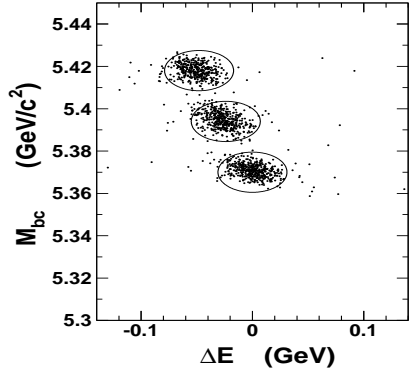}
\vspace{-0.2cm}
\caption{The $M_{\rm bc}$ and $\Delta E$ scatter plot
for the $B_s^0 \to D_s^- \pi^+$ decay obtained from
the MC simulation. The ellipses show the signal regions for the 
intermediate $B_s^* \bar{B}_s^*$ (top elliptical region), 
$B_s^* \bar{B}_s^0$ and $B_s^0 \bar{B}_s^*$ (middle elliptical region),
and $B_s^0 \bar{B}_s^0$ (bottom elliptical region) channels.}
\label{figamc}
\end{figure*}
\vspace{-0.3cm}

\section{RECENT MEASUREMENTS WITH 23.6\,fb{\boldmath $^{-1}$} AT THE {\boldmath $\Upsilon$}(5S)}
\subsection{Measurement of {\boldmath $B_s^0 \to D_s^- \pi^+$} decay and evidence for {\boldmath $B_s^0 \to D_s^{\mp} K^{\pm}$} decay}

We report here the preliminary results from studies of
$B_s^0 \to D_s^- \pi^+$ and $B_s^0 \to D_s^{\mp} K^{\pm}$ decays obtained
by the Belle collaboration with 23.6\,fb$^{-1}$ at the $\Upsilon$(5S)
\cite{bdspi}. In this analysis
$D_s^-$ candidates are reconstructed in the $\phi \pi^-$, $K^{*0} K^-$
and $K^0_S K^-$ modes.
Fig.\,3 shows $M_{\rm bc}$ and $\Delta E$ scatter plot for the
studied decays. A clear signal is observed in the $B_s^0 \to D_s^- \pi^+$
decay mode, and evidence for the $B_s^0 \to D_s^{\mp} K^{\pm}$ decay
is also seen.
For each mode, a two-dimensional unbinned extended maximum likelihood
fit in $M_{\rm bc}$ and $\Delta E$ is performed on the selected candidates.
Fig.\,4 shows the $M_{\rm bc}$ and $\Delta E$ projections
in the $B_s^* \bar{B}_s^*$ region of the data, together with the fitted
functions.

\vspace{-0.2cm}
\begin{figure*}[h]
\centering
\includegraphics[width=64mm]{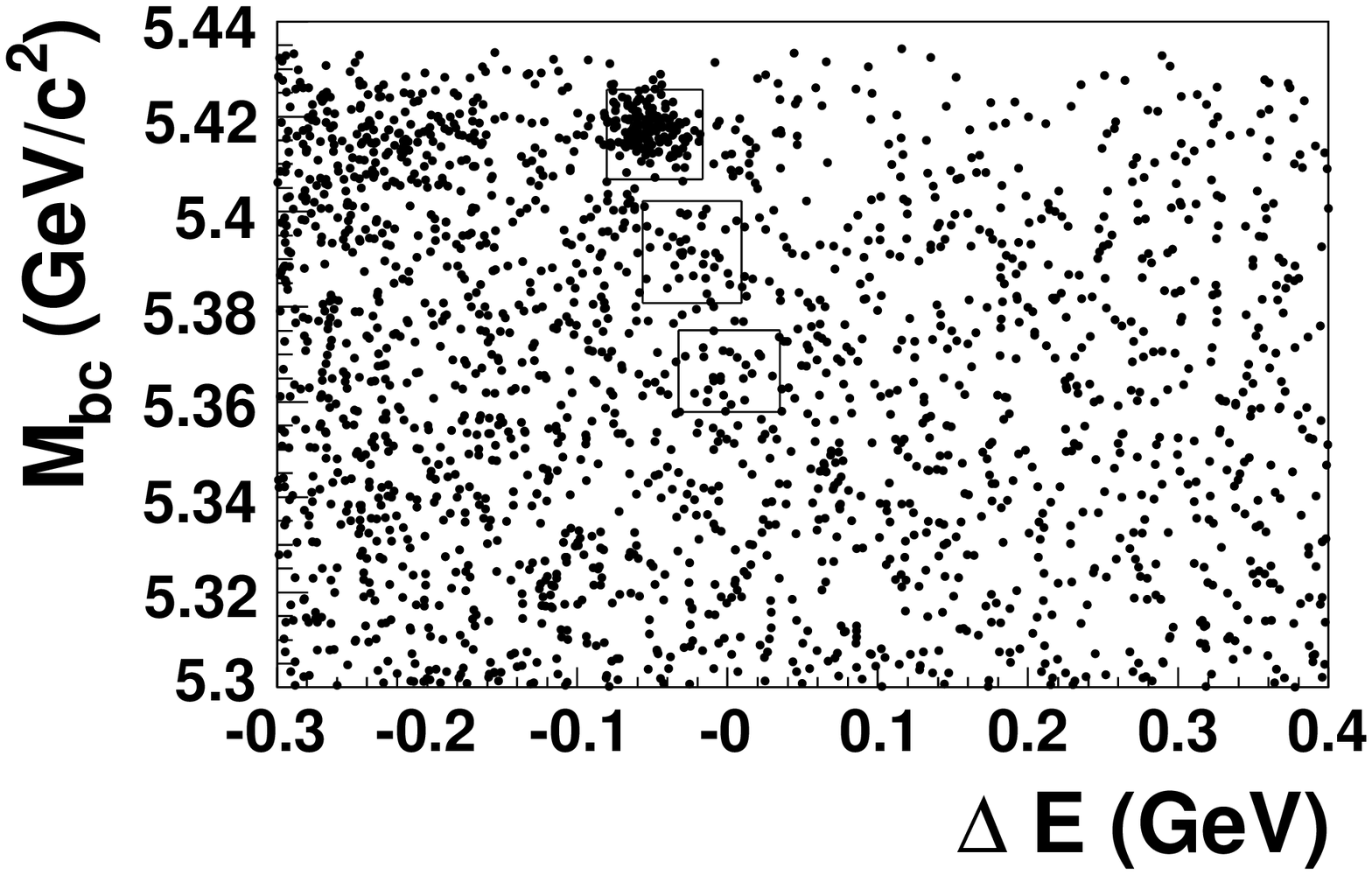}\hspace{0.5cm}\includegraphics[width=64mm]{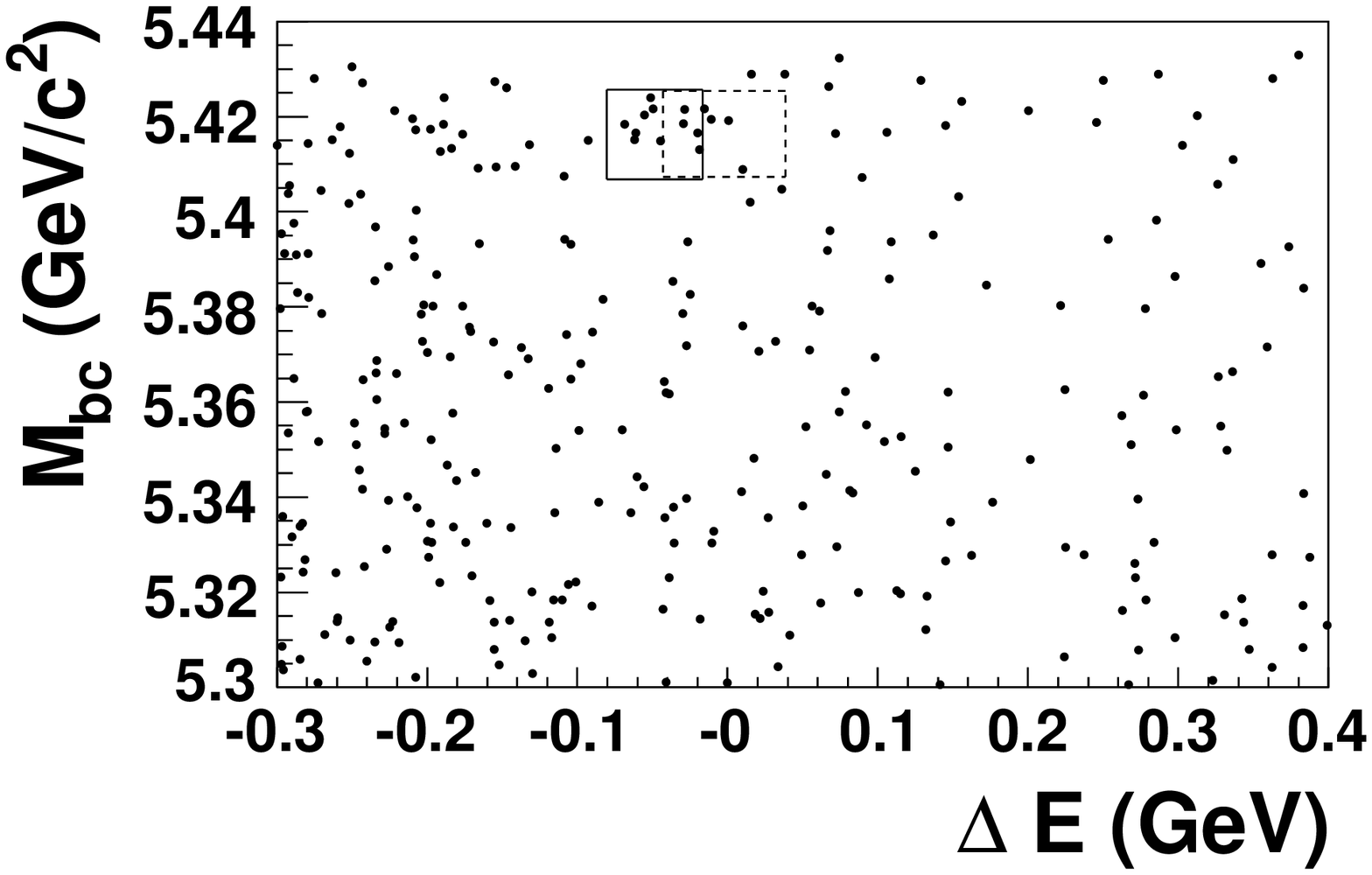}
\vspace{-0.2cm}
\caption{The $M_{\rm bc}$ and $\Delta E$ scatter plot for the
$B_s^0 \to D_s^- \pi^+$ (left) and $B_s^0 \to D_s^{\mp} K^{\pm}$ (right)
candidates. The three boxes in the left plot are the $\pm 2.5 \sigma$
signal regions ($B_s^* \bar{B}_s^*$,
$B_s^* \bar{B}_s^0$ and $B_s^0 \bar{B}_s^*$, from top to bottom)
while those in the right plot are the $\pm 2.5 \sigma$
$B_s^* \bar{B}_s^*$ regions for signal (solid) and for $B_s^0 \to D_s^- \pi^+$
background (dashed).}
\end{figure*}

\begin{figure*}[h]
\centering
\includegraphics[width=60mm]{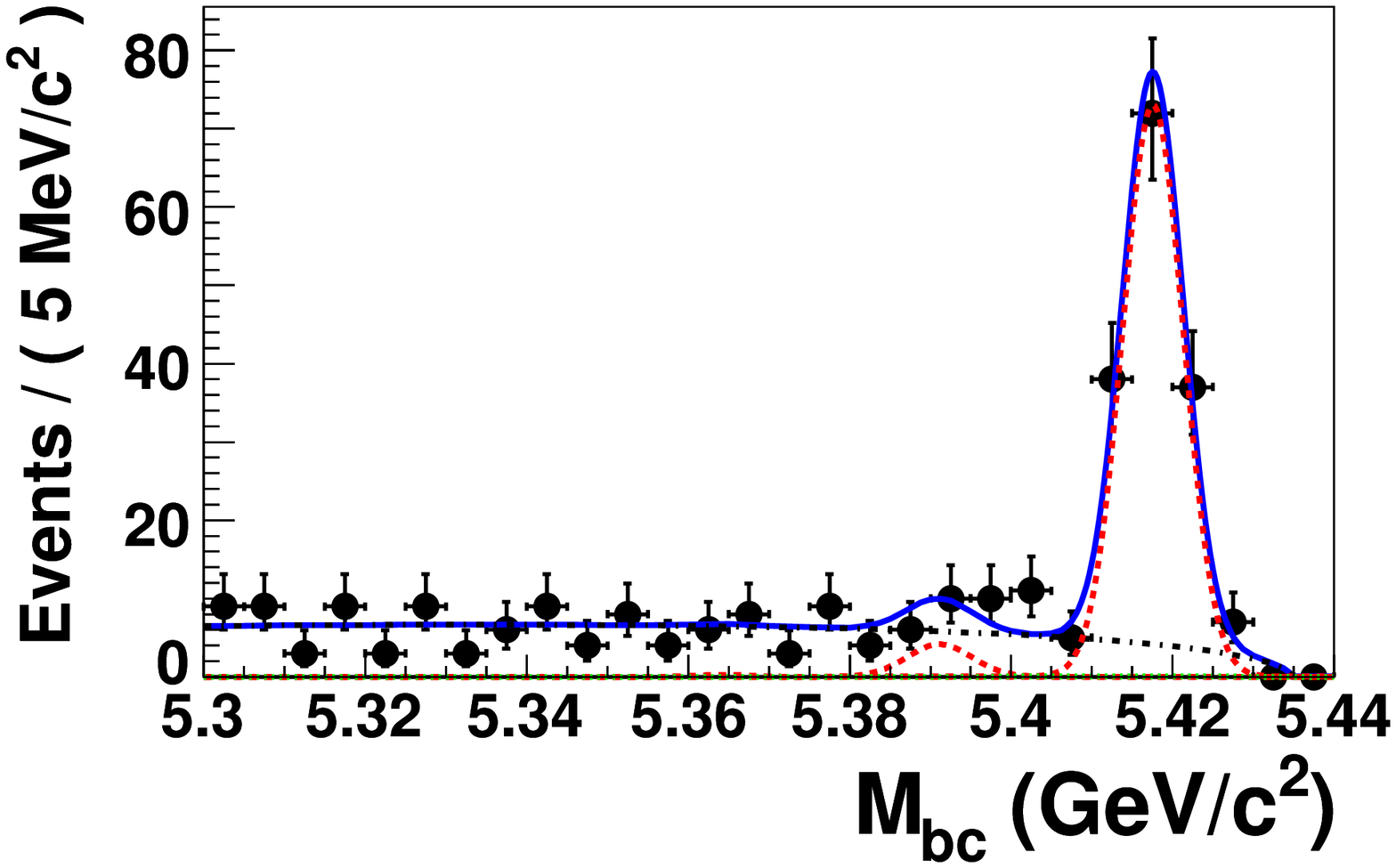}\hspace{0.5cm}\includegraphics[width=60mm]{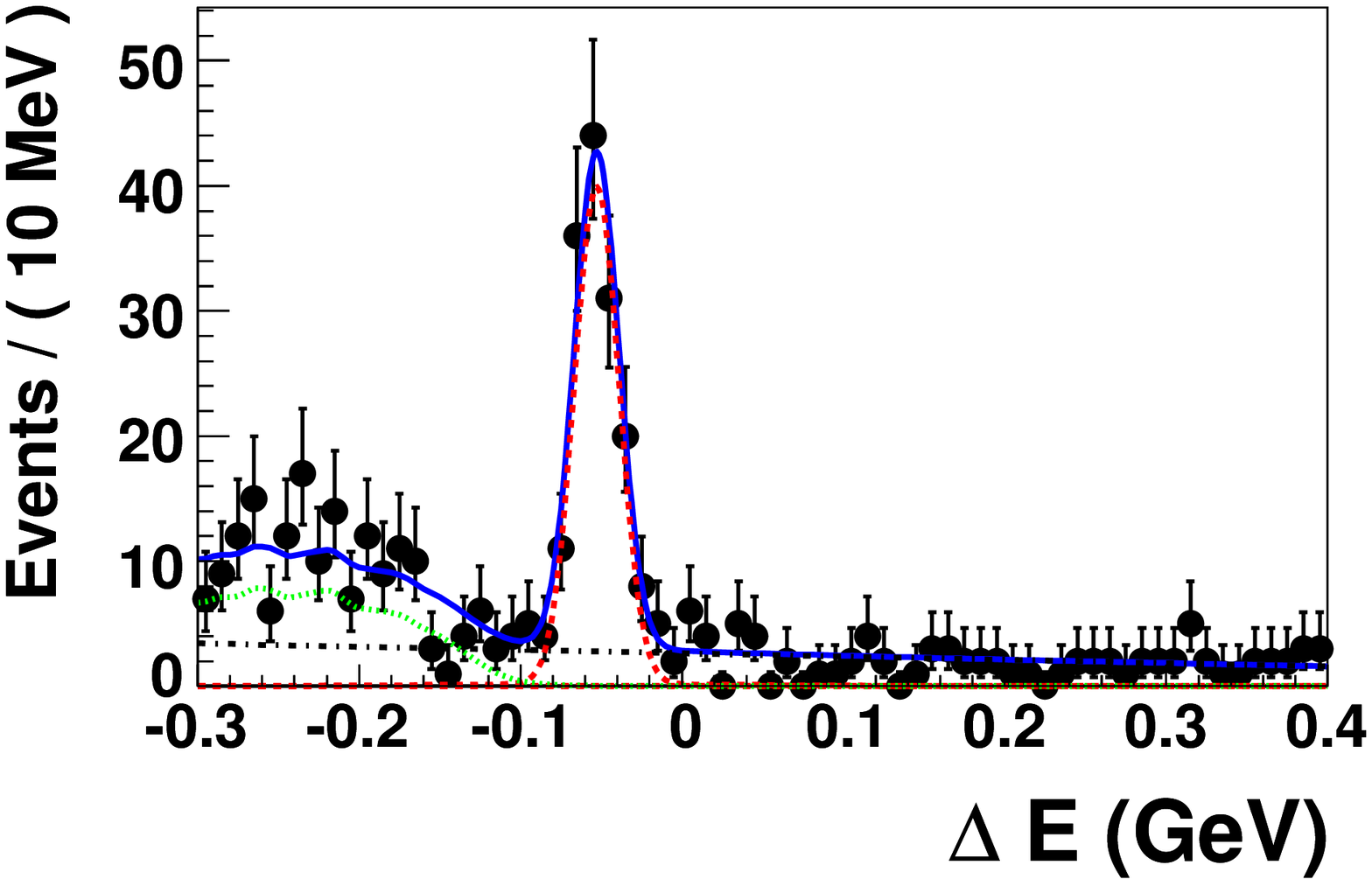}
\vspace{-0.1cm}
\caption{The $M_{\rm bc}$ distribution for $-80 < \Delta E < -17$ MeV region
(left) and $\Delta E$ distribution for $5.41 < M_{\rm bc} < 5.43\,$GeV/$c^2$
region (right) for the $B_s^0 \to D_s^- \pi^+$ candidates.
The different fitted components are shown with dashed curves for the
signal, dotted curves for the $B_s^0 \to D_s^{*-} \pi^+$ background,
and dash-dotted curves for the continuum.}
\end{figure*}

Finally, the branching fractions 
${\cal B}(B_s^0 \to D_s^- \pi^+) = (3.33^{+0.32}_{-0.30}{\rm (stat.)} ^{+0.41}_{-0.40}{\rm (syst.)} \pm 0.44 (f_s) ^{+0.33}_{-0.27}({\cal B}(D_s^- \to \phi \pi^-))) \times 10^{-3}$ 
and 
${\cal B}(B_s^0 \to D_s^{\mp} K^{\pm}) = (2.2^{+1.1}_{-0.9}{\rm (stat.)} \pm 0.3 {\rm (syst.)} \pm 0.3 (f_s) \pm 0.2 ({\cal B}(D_s^- \to \phi \pi^-))) \times 10^{-4}$
are measured.
The ratio ${\cal B}(B_s^0 \to D_s^{\mp} K^{\pm}) / {\cal B}(B_s^0 \to D_s^- \pi^+) = (6.5^{+3.5}_{-2.9})\%$ is derived; the errors are completely
dominated by the low $B_s^0 \to D_s^{\mp} K^{\pm}$ statistics.

Comparing the number of events reconstructed in the 
$B_s^0 \to D_s^- \pi^+$ mode in three signal regions, 
the fraction of $B_s^\ast\bar{B}_s^\ast$
events over all $B_s^{(\ast)}\bar{B}_s^{(\ast)}$ events
was measured to be
$f_{B_s^\ast\bar{B}_s^\ast}=(90.3^{+3.8}_{-4.0})\%$.
This number is higher than the value of 70\,$\%$ predicted
by several theoretical models \cite{tdspia,tdspib}.
From the $B_s^0$ signal fit the masses 
$m(B_s^*) = (5317.6 \pm 0.4 \pm 0.5)\,$MeV/$c^2$ and 
$m(B_s^0) = (5364.6 \pm 1.3 \pm 2.4)\,$MeV/$c^2$ are obtained.
In contrast to theoretical predictions \cite{tdspic}
the mass difference $m(B_s^*)-m(B_s^0)$ 
obtained is 2.7$\sigma$ larger than the world average for
$m(B^{*0})-m(B^0)$.

The distribution of the cosine of the angle between the
$B_s^0$ momentum and the beam axis in the CM system for the
$\Upsilon$(5S)$\to B_s^* \bar{B}_s^*$ decay 
(in the $B_s^0 \to D_s^- \pi^+$ mode) is shown in Fig.~5.
The flat shape of this distribution
indicates that different spin-momentum combinations
contribute in the $\Upsilon$(5S)$\to B_s^* \bar{B}_s^*$ decay.

\vspace{-0.1cm}
\begin{figure*}[h]
\centering
\includegraphics[width=59mm]{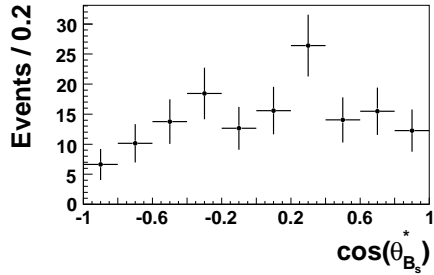}
\vspace{-0.3cm}
\caption{The cosine of the angle between the 
$B_s^0$ momentum and the beam axis in the CM system for the
$\Upsilon$(5S)$\to B_s^* \bar{B}_s^*$ decay.}
\end{figure*}

\vspace{-0.2cm}
\subsection{Observation of {\boldmath $B_s^0 \to \phi \gamma$} and search for {\boldmath $B_s^0 \to \gamma \gamma$} decays}

The radiative decays $B_s^0 \to \phi \gamma$ and $B_s^0 \to \gamma \gamma$
have been studied at the $\Upsilon$(5S) with 23.6\,fb$^{-1}$ \cite{bfiga}.
Only upper limits were obtained for these decays in the previous Belle 
analysis \cite{bele}, based on 1.86\,fb$^{-1}$ of data.

Within the Standard Model (SM) the $B_s^0 \to \phi \gamma$ decay 
can be described
by a radiative penguin diagram (Fig.~6, left) and the
corresponding branching fraction is predicted 
to be $\sim4 \times 10^{-5}$ \cite{tfigaa}.
The $B_s^0 \to \gamma \gamma$ decay is expected to proceed via a
penguin annihilation diagram (Fig.~6, right) and to have a branching
fraction in the range $(0.5-1.0)\times 10^{-6}$. 
However, the $B_s^0 \to \gamma \gamma$ decay branching fraction
is sensitive to some BSM contributions and can be enhanced by about an order 
of magnitude \cite{tfigab,tfigac,tfigad};
such enhanced values are not far from the upper limit expected 
in this analysis.

\begin{figure*}[h]
\centering
\includegraphics[width=39mm]{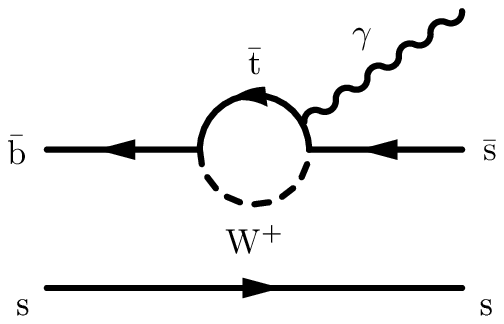}\hspace{2cm}\includegraphics[width=39mm]{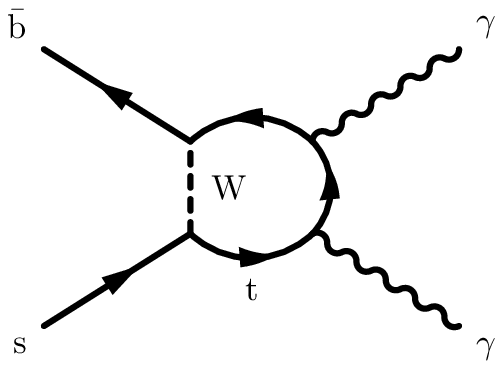}
\vspace{-0.1cm}
\caption{Diagrams describing the dominant SM processes for the
$B_s^0 \to \phi \gamma$ (left) and $B_s^0 \to \gamma \gamma$ (right) decays.}
\end{figure*}

The three-dimensional (two-dimensional) unbinned extended 
maximum likelihood fit to $M_{\rm bc}$, $\Delta E$
and cos$\theta_\phi^h$ ($M_{\rm bc}$ and $\Delta E$)
is performed for $B_s^0 \to \phi \gamma$ ($B_s^0 \to \gamma \gamma$)
decay to extract the signal yield.
Fig.~7 shows the $M_{\rm bc}$ and $\Delta E$ projections
of the data, together with the fitted functions.

A clear signal is seen in the $B_s^0 \to \phi \gamma$ mode.
This radiative decay is observed for the
first time and the branching fraction 
${\cal B} (B_s^0 \to \phi \gamma) = (5.7^{+1.8}_{-1.5} {\rm (stat.)} ^{+1.2}_{-1.1} {\rm (syst.)}) \times 10^{-5}$ 
is measured. The obtained value is in agreement
with the SM predictions. No significant signal is observed in the
$B_s^0 \to \gamma \gamma$ mode, and an upper limit at the $90\%$ C.L.
of ${\cal B} (B_s^0 \to \phi \gamma) < 8.7 \times 10^{-6}$ is set.

\begin{figure*}[h]
\centering
\includegraphics[width=50mm]{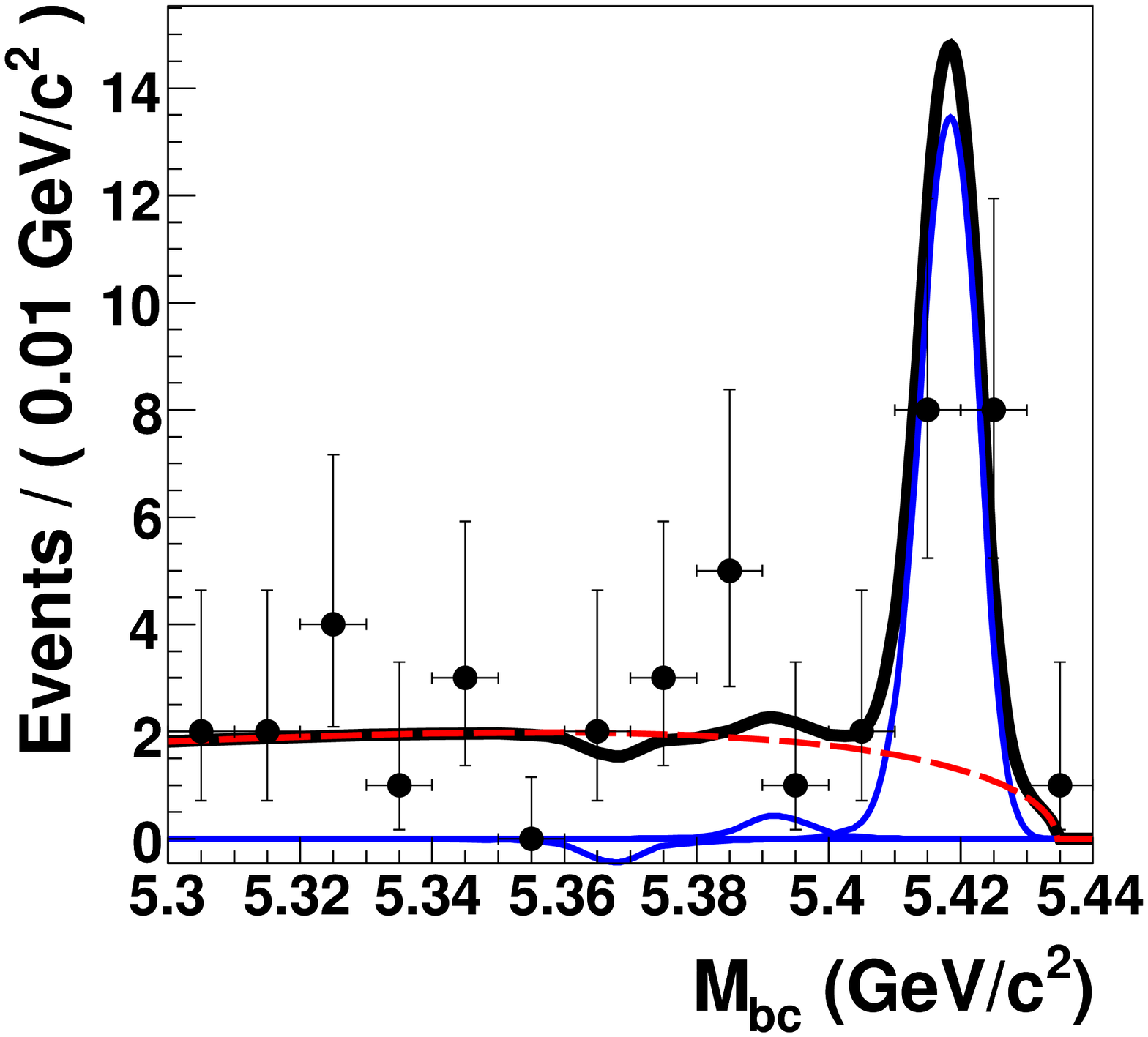}\hspace{0.5cm}\includegraphics[width=50mm]{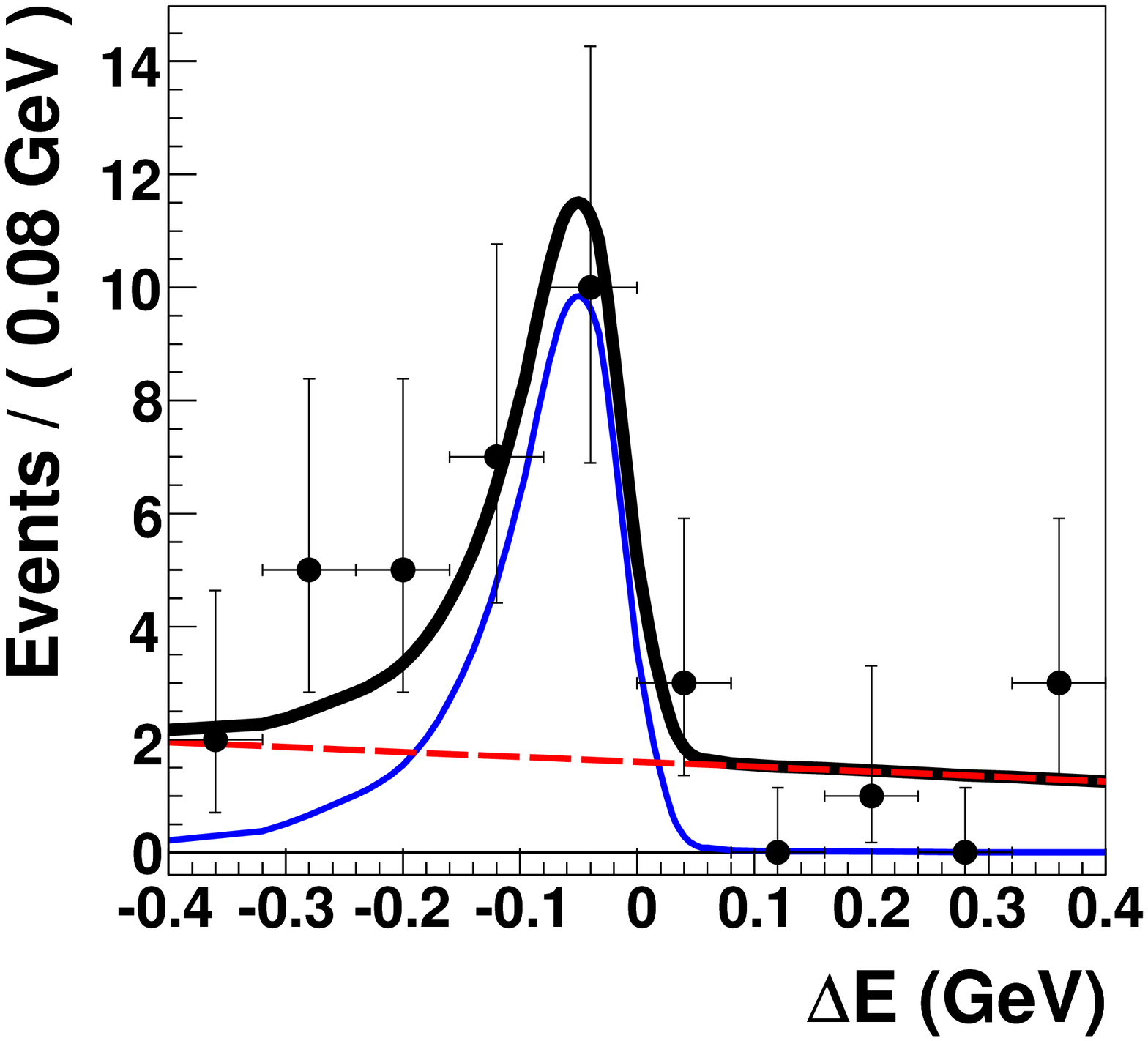} \\
\includegraphics[width=50mm]{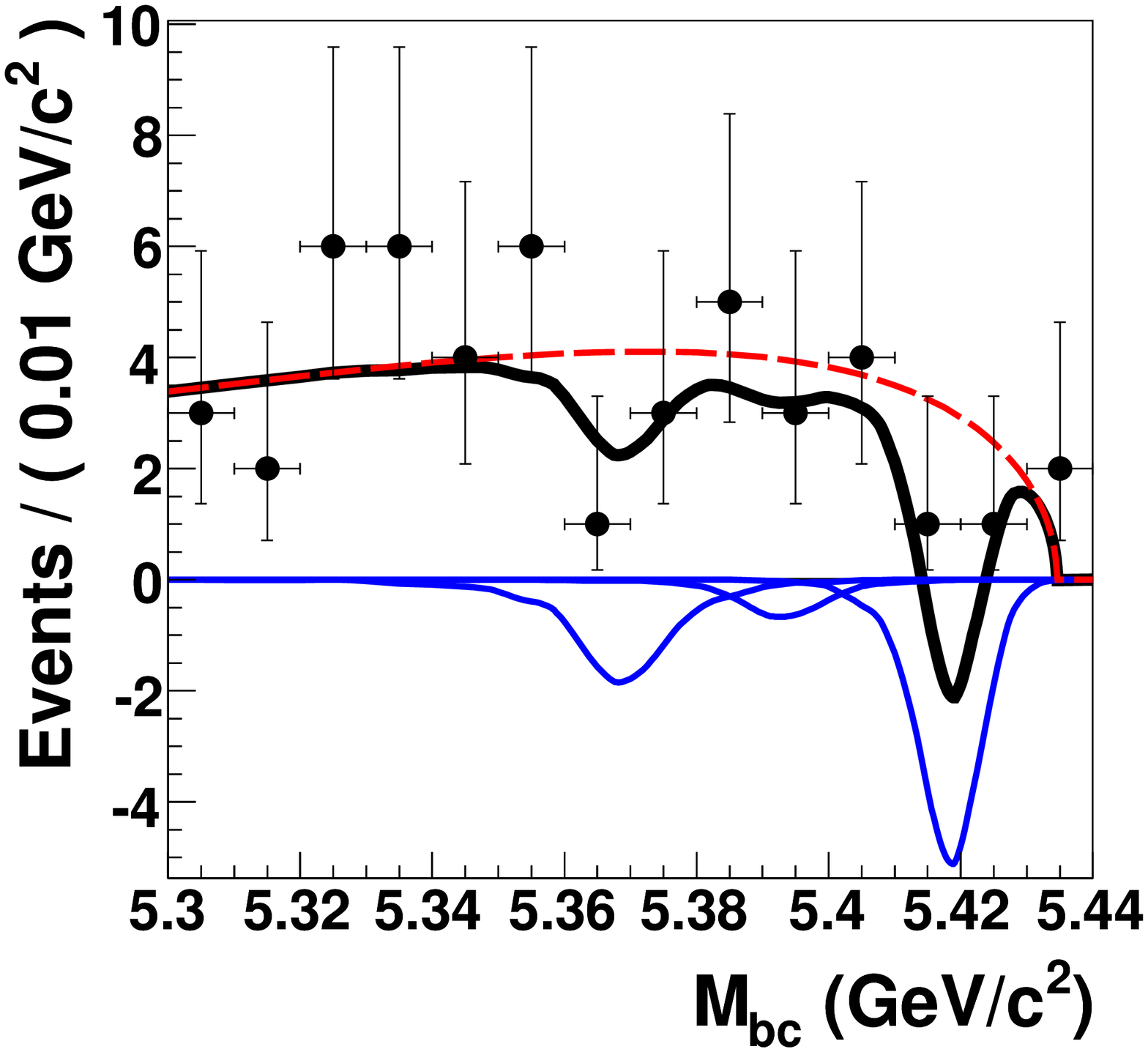}\hspace{0.5cm}\includegraphics[width=50mm]{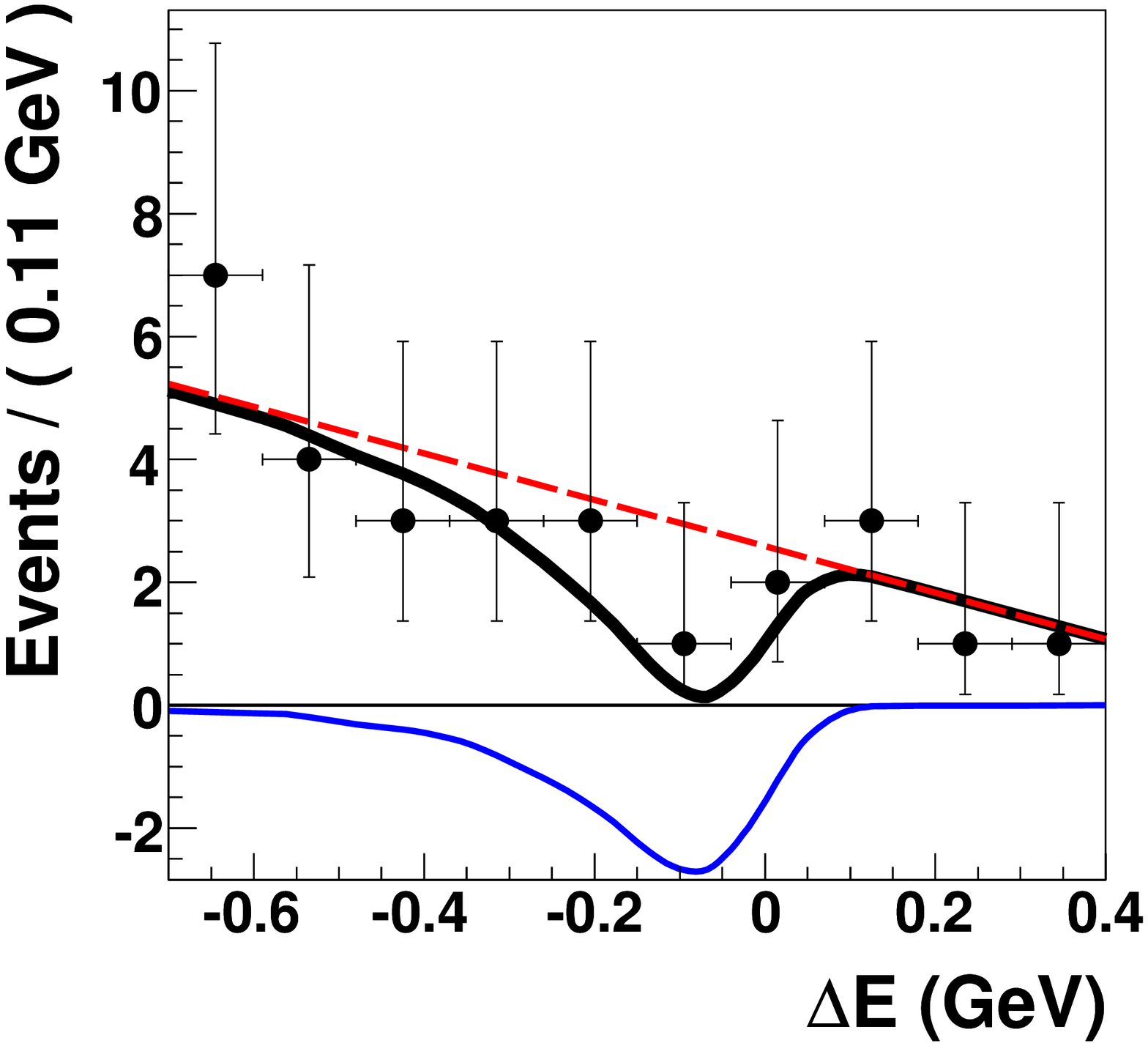}
\vspace{-0.1cm}
\caption{The $M_{\rm bc}$ projection (left) and $\Delta E$ projection 
(right) for the $B_s^0 \to \phi \gamma$ (top) and $B_s^0 \to \gamma \gamma$ 
(bottom) modes.
The points with error bars represent data, the thick solid curves are 
the fit functions, the thin solid curves are the signal functions, 
and the dashed curves show the continuum contribution.
On the $M_{\rm bc}$ figure, signals from $B_s^0 \bar{B}_s^0$,
$B_s^* \bar{B}_s^0$, and $B_s^* \bar{B}_s^*$ appear from left to right.
On the $\Delta E$ figure, due to the requirement $M_{\rm bc} > 5.4\,$GeV/$c^2$
only the $B_s^* \bar{B}_s^*$ signal contributes.}
\end{figure*}

\subsection{First measurement of {\boldmath $B_s^0 \to X^+ \ell^- \nu$} decay}

The inclusive semileptonic $B_s^0 \to X^+ \ell^- \nu$ decay branching
fraction has been measured
for the first time using a 23.6\,fb$^{-1}$ data sample collected
at the $\Upsilon$(5S) resonance \cite{bdsle}.
This measurement is of special interest, because
the total semileptonic $B^0$, $B^+$ and $B_s^0$ branching 
fractions, together with corresponding well-measured lifetimes,
determine the semileptonic widths.
These widths for the $B^0$, $B^+$ and $B_s^0$ mesons are expected to be
equal, neglecting small corrections due to electromagnetic and
light quark mass difference effects.
If any significant difference between the $B^0$, $B^+$ and $B_s^0$
semileptonic widths were observed, it would indicate
a previously unknown source of lepton production in $B$ decays.
On the other hand, assuming equal semileptonic widths in $B^{0/+}$ and $B_s^0$
decays, the inclusive $B_s^0$ semileptonic branching fraction can be directly 
related to the $B_s^0$ lifetime. This can be potentially useful
taking into account a possible $CP$ correlation of the $B_s^0$ mesons produced
in $\Upsilon$(5S) decays.

The correlated production of a $D_s^+$ meson and a same-sign
lepton at the $\Upsilon$(5S) resonance is used in this analysis 
to measure ${\cal B}(B_s^0 \rightarrow X^+ \ell^- \nu)$.
$D_s^+$ candidates are reconstructed in a clean $\phi \pi^+$ mode.
First, we selected the data sample with $D_s^+$ mesons and then
studied the same-sign lepton production in this sample.
Neither the $c\bar{c}$ continuum nor $B^{(*)}\bar{B}^{(*)}$ states
(except for a small contribution due to $\sim 19\%$ $B^0$ mixing effect) 
can result in a same-sign $c$-quark (i.e.,\ $D_s^+$ meson) and 
primary lepton final state. 

The final leptonic momentum distributions produced 
in $B_s^0$ decays (Fig. 8)
obtained after background subtractions and an efficiency correction
are used to extract the numbers of primary and secondary leptons.
The backgrounds are subtracted using continuum and $\Upsilon$(4S)
data to estimate continuum and $B\bar{B}$ contributions. 
The MC simulation is used to obtain the momentum distribution shapes
for leptons from primary $B_s$ decays
and from secondary $D_s^+$, $D^0$, $D^+$ or $\tau^+$ decays.
We fit the data with a function which includes the sum of
these two terms with fixed shapes and floating normalizations.

\begin{figure*}[h]
\centering
\includegraphics[width=45mm,height=40mm]{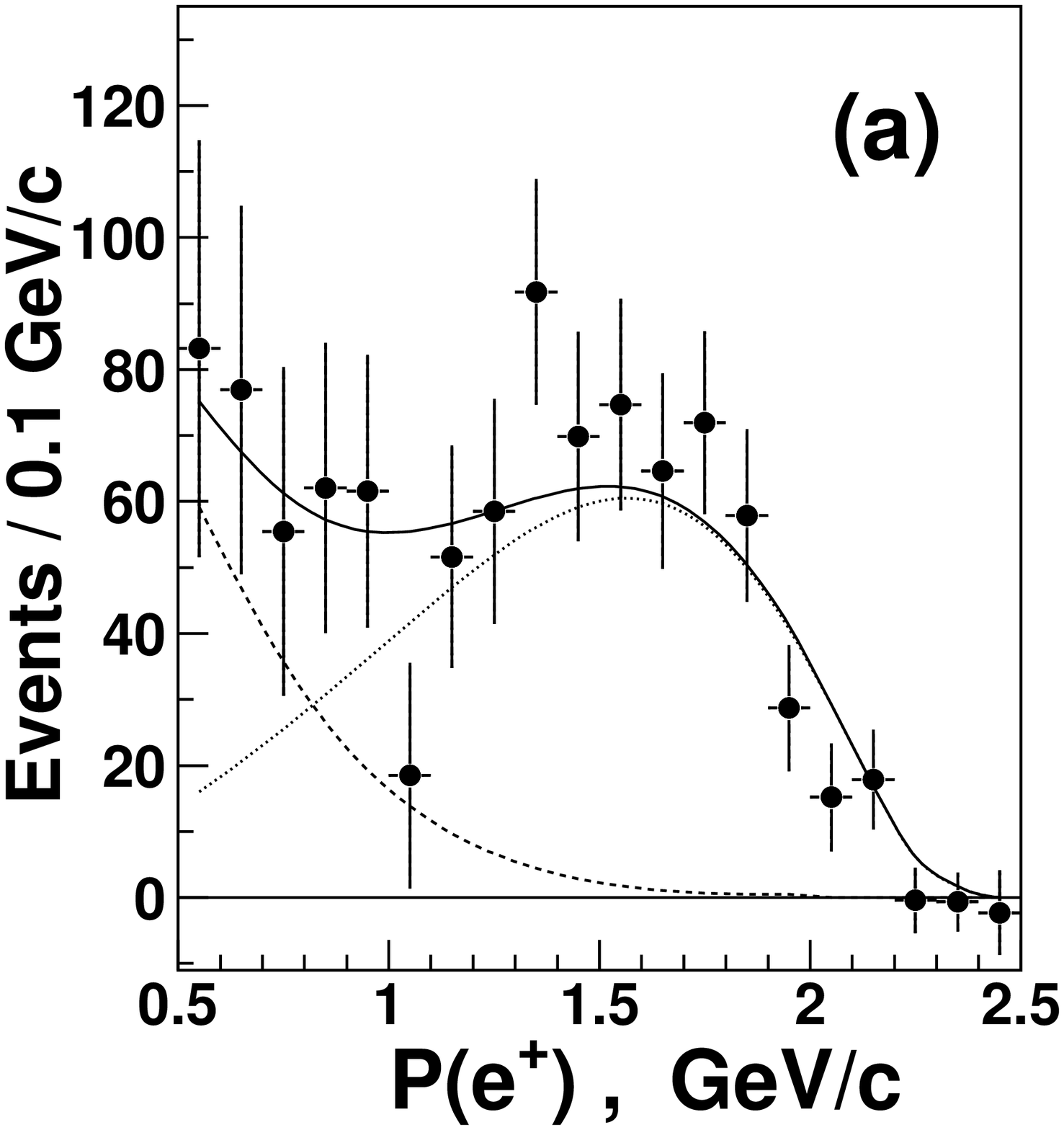}\hspace{0.7cm}\includegraphics[width=45mm,height=40mm]{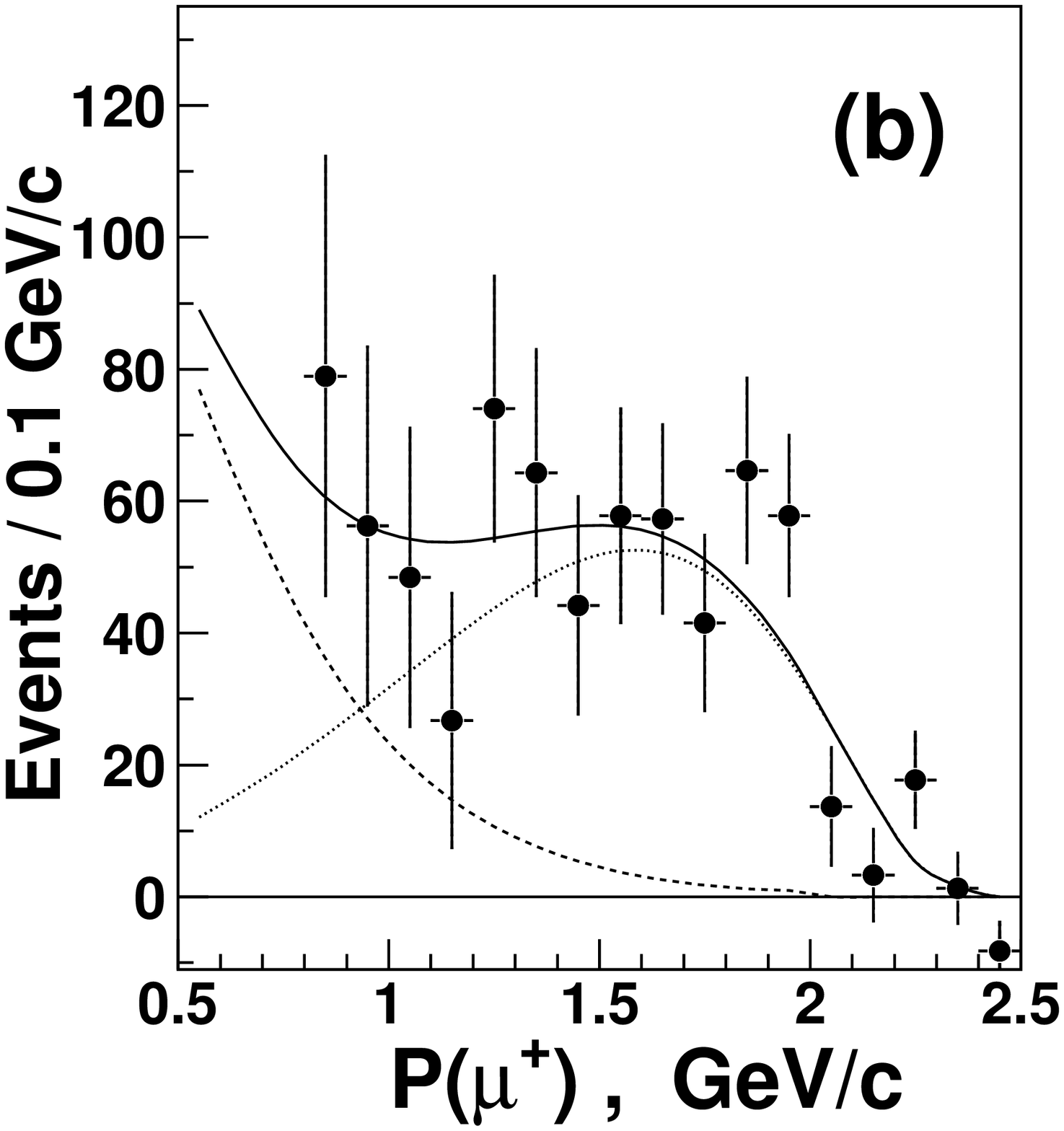}
\vspace{-0.1cm}
\caption{The electron (a) and muon (b)
momentum distributions from $B_s^0$ decays. 
The solid curves show the results of the fits, and the dotted 
curves show the fitted contributions from primary and secondary leptons.}
\label{figc}
\end{figure*}

Finally, we obtained the semileptonic branching fractions:
\begin{eqnarray}
\nonumber
{\cal B}(B_s^0 \rightarrow X^+ e^- \nu)\, = (10.9 \pm 1.0 \pm 0.9)\% \\
{\cal B}(B_s^0 \rightarrow X^+ \mu^- \nu)\, = (9.2 \pm 1.0 \pm 0.8) \% \\
\nonumber
{\cal B}(B_s^0 \rightarrow X^+ \ell^- \nu)\, = (10.2 \pm 0.8 \pm 0.9)\% ,
\end{eqnarray}
\noindent
where the latter one represents an average over electrons and muons.
The obtained branching fractions can be compared with the PDG value
${\cal B}(B^0 \rightarrow X^+ \ell^- \nu)\, = (10.33 \pm 0.28)\%$ \cite{pdg},
which is theoretically expected to be approximately the same,
neglecting a small possible lifetime difference and
small corrections due to electromagnetic and
light quark mass difference effects.

\subsection{Observation of {\boldmath $\Upsilon{\rm (5S)} \to \Upsilon {\rm (1S)}\, \pi^+\pi^-$} and {\boldmath $\Upsilon{\rm (5S)} \to \Upsilon{\rm (2S)}\, \pi^+\pi^-$} decays}

The production of $\Upsilon{\rm (1S)}\, \pi^+\pi^-$, 
$\Upsilon{\rm (2S)}\, \pi^+\pi^-$, $\Upsilon{\rm (3S)}\, \pi^+\pi^-$,
and $\Upsilon{\rm (1S)}\, K^+K^-$ final states in a 21.7\,fb$^{-1}$
data sample obtained at $e^+ e^-$ collisions with CM energy near the peak 
of the $\Upsilon$(5S) resonance has been studied by Belle \cite{bups}.
Final states with two opposite-sign muons and two opposite-sign
pions (or kaons) are selected. 
The signal candidates are identified using the kinematic variable
$\Delta M$, defined as the difference between $M(\mu^+\mu^-\pi^+\pi^-)$
or $M(\mu^+\mu^-K^+K^-)$ and $M(\mu^+\mu^-)$ for pion or kaon modes.
In the studied processes the opposite-sign muon pair should have a mass in 
the nominal $\Upsilon$(1S), $\Upsilon$(2S) or $\Upsilon$(3S) mass regions.

Figure 9 (left) shows the two-dimensional scatter plots of $M(\mu^+\mu^-)$
vs $\Delta M$. Horizontal shaded bands correspond
to $\Upsilon$(1S), $\Upsilon$(2S) or $\Upsilon$(3S)
(only $\Upsilon$(1S) for (b)), and open boxes are the fitting
regions for $\Upsilon{\rm (5S)} \to \Upsilon{\rm (nS)}\, \pi^+\pi^-$
and $\Upsilon{\rm (5S)} \to \Upsilon{\rm (1S)}\, K^+K^-$.
The region where the mass of $\mu^+\mu^-\pi^+\pi^-$ (or $\mu^+\mu^-K^+K^-$)
combination corresponds to the $\Upsilon$(5S) CM energy of 10869 MeV
is tilted in the $M(\mu^+\mu^-)$ vs $\Delta M$ plot.
Fig.~9 (right) shows the $\Delta M$ projections for
$\mu^+\mu^-\pi^+\pi^-$ events in the $\Upsilon$(1S) (a) and
$\Upsilon$(2S) (b) regions.

\begin{figure*}[h]
\centering
\includegraphics[width=80mm,height=50mm]{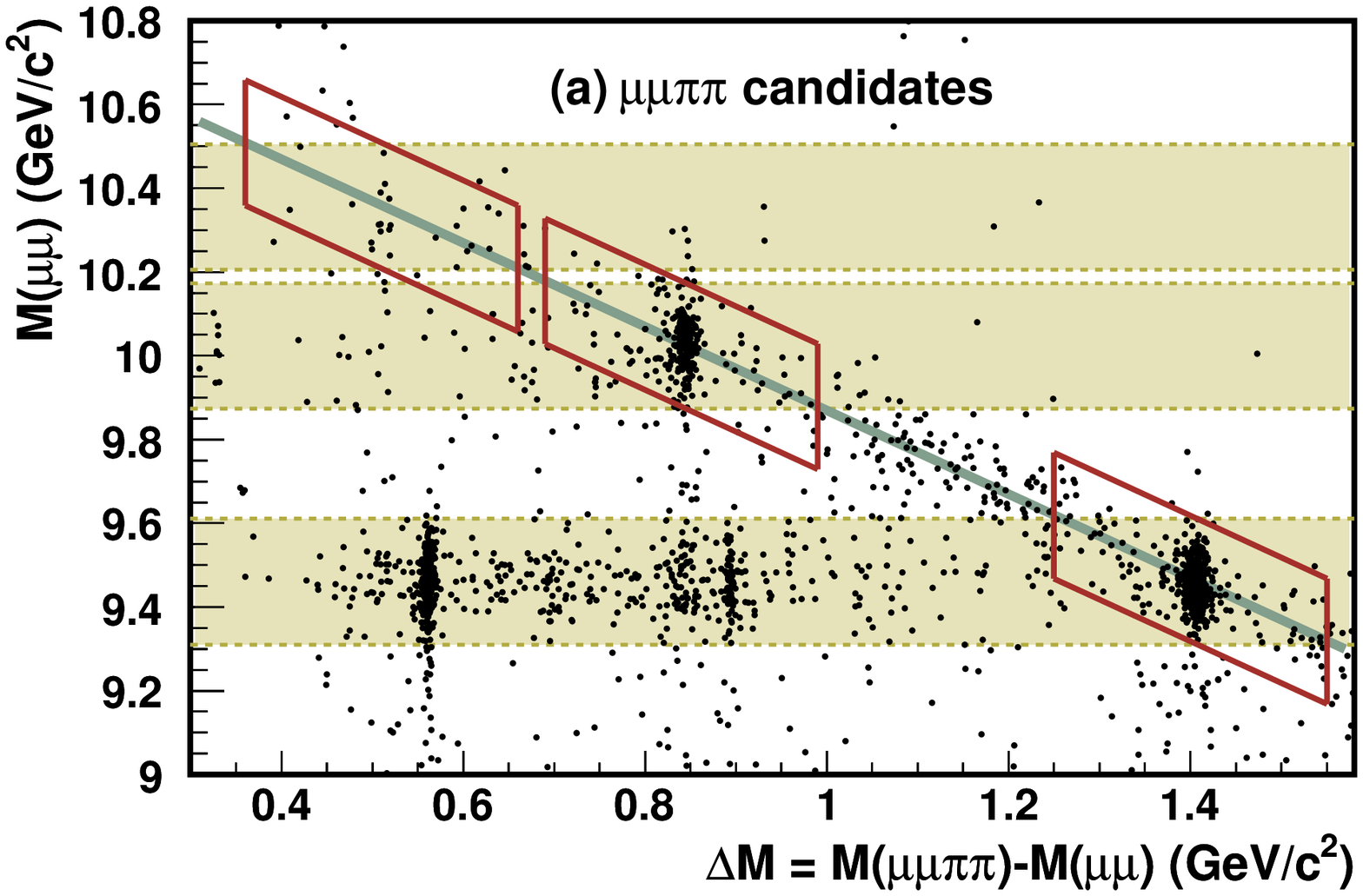}\hspace{0.7cm}\includegraphics[width=80mm,height=30mm]{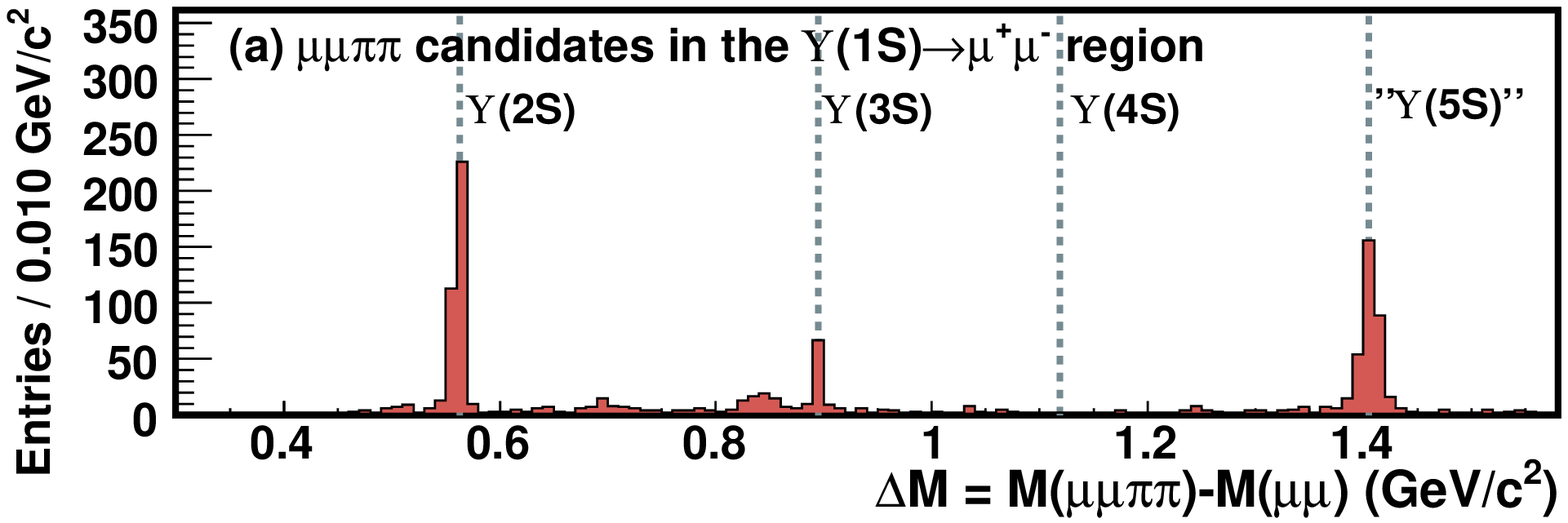}
\includegraphics[width=80mm,height=30mm]{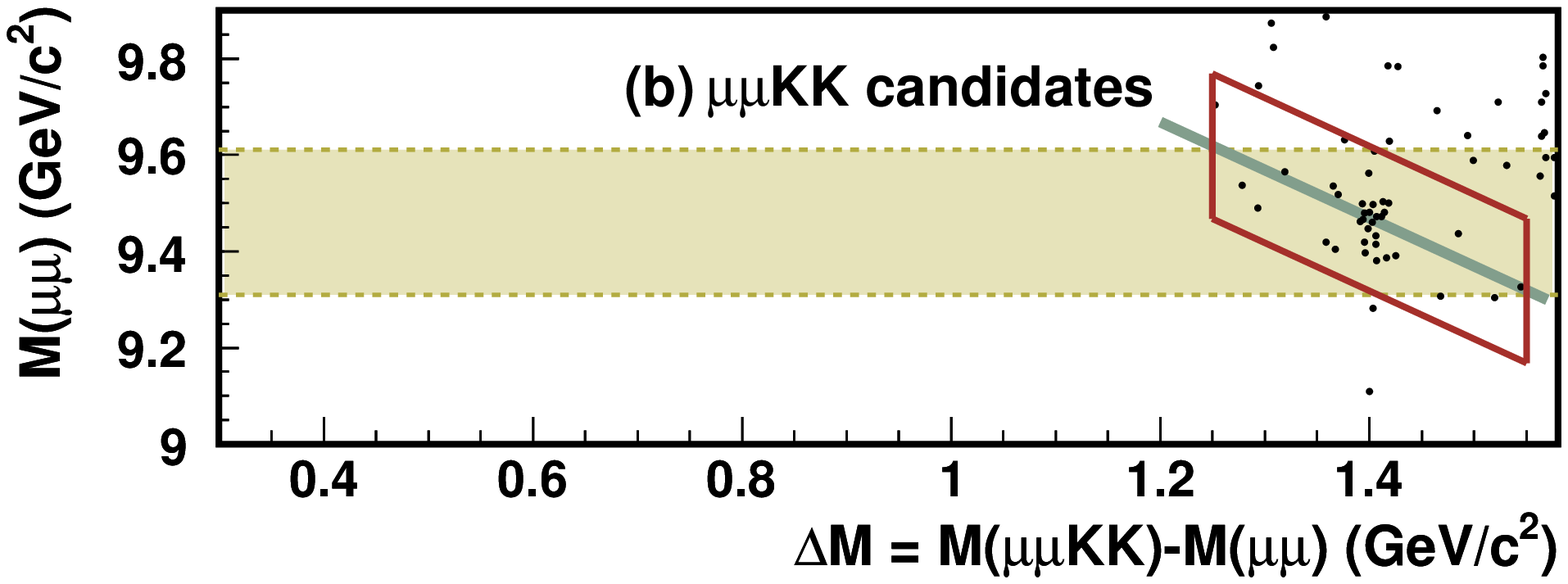}\hspace{0.7cm}\includegraphics[width=80mm,height=30mm]{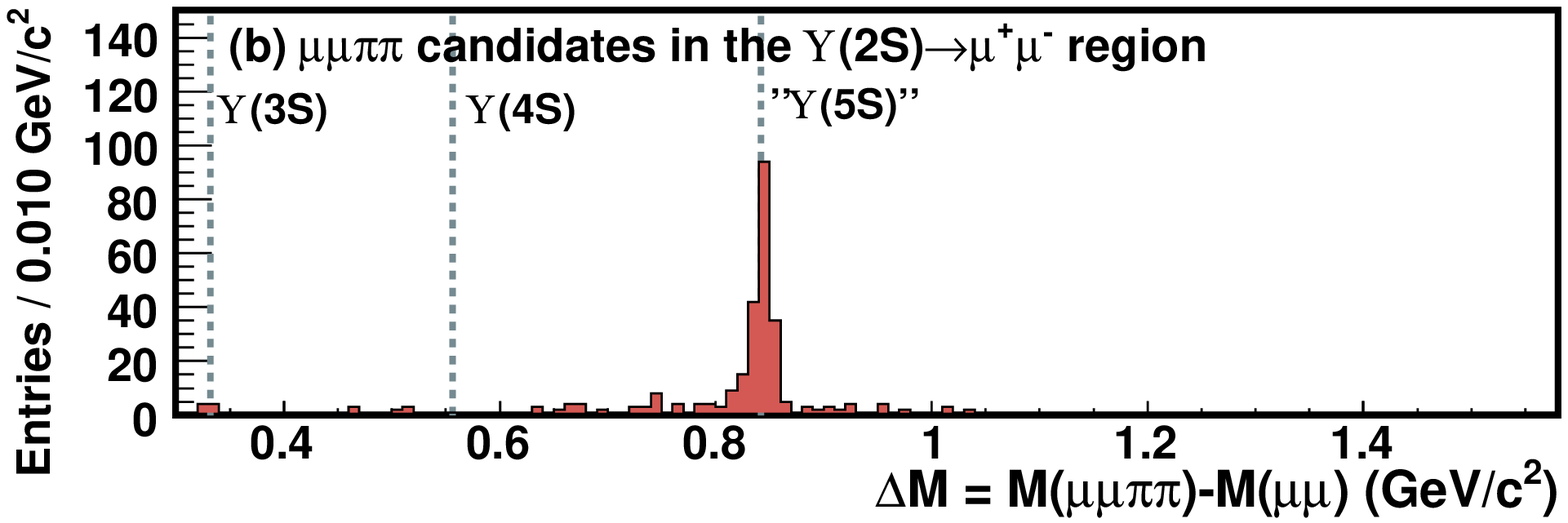}
\vspace{-0.1cm}
\caption{Scatter plot of $M(\mu^+\mu^-)$ vs $\Delta M$ for the data
collected at $\sqrt{s} \sim 10.87\,$GeV, for (a, left) $\mu^+\mu^-\pi^+\pi^-$
and (b, left) $\mu^+\mu^-K^+K^-$ candidates. Horizontal shaded bands correspond
to $\Upsilon$(1S), $\Upsilon$(2S) or $\Upsilon$(3S)
(only $\Upsilon$(1S) for (b)), and open boxes are the fitting
regions for $\Upsilon{\rm (5S)} \to \Upsilon{\rm (nS)}\, \pi^+\pi^-$
and $\Upsilon{\rm (5S)} \to \Upsilon{\rm (1S)}\, K^+K^-$.
The $\Delta M$ projections in the $\Upsilon$(1S) (a, right) and 
$\Upsilon$(2S) (b, right) regions are also shown. Vertical dashed
lines show the expected $\Delta M$ values for the 
$\Upsilon{\rm (nS)} \to \Upsilon{\rm (1,2S)}\, \pi^+\pi^-$ transitions.}
 
\label{figc}
\end{figure*}

The obtained branching fractions and partial widths are
given in Table 1. For comparison, the partial widths
for similar transitions from $\Upsilon$(2S), $\Upsilon$(3S), or $\Upsilon$(4S)
are also shown. The $\Upsilon$(5S) partial widths
(assuming that the signal events are solely due to the $\Upsilon$(5S)
resonance) are found to be in the range (0.52-0.85)\,MeV, that is
more than 2 orders of magnitude larger than the corresponding
partial widths for $\Upsilon$(2S), $\Upsilon$(3S), or $\Upsilon$(4S) decays. 
The unexpectedly large $\Upsilon$(5S) partial widths disagree with 
the expectation for a pure $b\bar{b}$ state, unless there is a new mechanism
to enhance the decay rates.

\begin{table}[h]
\begin{center}
\caption{The branching fractions (${\cal B}$) and the partial widths
($\Gamma$) for $\Upsilon{\rm (nS)} \to \Upsilon{\rm (mS)}\, \pi^+\pi^-$
and $\Upsilon{\rm (1S)}\, K^+K^-$ processes are listed. The first error is
statistical, and the second is systematic. The values for the
$\Upsilon$(2,3,4S) decays are from \cite{pdg}.}\vspace{0.1cm}
\begin{tabular}
{|@{\hspace{0.3cm}}l@{\hspace{0.3cm}}|@{\hspace{0.3cm}}c@{\hspace{0.3cm}}|@{\hspace{0.3cm}}c@{\hspace{0.3cm}}|@{\hspace{0.3cm}}l@{\hspace{0.3cm}}|@{\hspace{0.3cm}}c@{\hspace{0.3cm}}|}
\hline \textbf{Process} & \textbf{${\cal B}$ ($\%$)} & \textbf{$\Gamma$ (MeV)} & 
\textbf{Process} & \textbf{$\Gamma$ (MeV)}
\\
\hline 
$\Upsilon{\rm (5S)} \to \Upsilon{\rm (1S)}\, \pi^+\pi^-$ & $0.53 \pm 0.03 \pm 0.05$ & $0.59 \pm 0.04 \pm 0.09$ & $\Upsilon{\rm (2S)} \to \Upsilon{\rm (1S)}\, \pi^+\pi^-$ & 0.006 \\
$\Upsilon{\rm (5S)} \to \Upsilon{\rm (2S)}\, \pi^+\pi^-$ & $0.78 \pm 0.06 \pm 0.11$ & $0.85 \pm 0.07 \pm 0.16$ & $\Upsilon{\rm (3S)} \to \Upsilon{\rm (1S)}\, \pi^+\pi^-$ & 0.0009 \\
$\Upsilon{\rm (5S)} \to \Upsilon{\rm (3S)}\, \pi^+\pi^-$ & $0.48^{+0.18}_{-0.15} \pm 0.07$ & $0.52^{+0.20}_{-0.17} \pm 0.10$ & $\Upsilon{\rm (4S)} \to \Upsilon{\rm (1S)}\, \pi^+\pi^-$ & 0.0019 \\
$\Upsilon{\rm (5S)} \to \Upsilon{\rm (1S)}\, K^+K^-$ & $0.061^{+0.016}_{-0.014} \pm 0.010$ & $0.67^{+0.017}_{-0.015} \pm 0.013$ & & \\
\hline
\end{tabular}
\label{l2ea4-t1}
\end{center}
\end{table}

\section{PHYSICS PROSPECTS AT THE {\boldmath $\Upsilon$}(5S)}

\subsection{Physics with (20-200)\,fb{\boldmath $^{-1}$} at the {\boldmath $\Upsilon$}(5S)}

Data at the $\Upsilon$(5S) have many advantages 
for $B_s^0$ studies compared to hadron-hadron collisions
(in particular with the CDF and D0 experiments at Tevatron),
such as high photon and $\pi^0$ reconstruction efficiency,
trigger efficiency of almost 100$\%$ for hadronic modes and
excellent charged kaon and pion identification.
A model-independent determination of the number of initial $B_s^0$ mesons
at $\Upsilon$(5S) data samples opens an opportunity 
for precise absolute $B_s^0$ branching fraction measurements.
The possibility of partial reconstruction of specific $B_s^0$ decays
and measurements of inclusive $B_s^0$ processes
are additional advantages of $e^+ e^-$ colliders running
at the $\Upsilon$(5S).
On the other hand, lower $B^0$ and $B^+$ meson production rates
at the $\Upsilon$(5S) compared to that at the $\Upsilon$(4S),
the smaller number of produced $B_s^0$ mesons compared to the Tevatron
experiments, and insufficient accuracy of the $B_s^0$ vertex 
reconstruction to observe $B_s^0-\bar{B}_s^0$ mixing
are the main disadvantages of running at the $\Upsilon$(5S).

Developing our current and future physics programs at the $\Upsilon$(5S)
is based on the advantages.
The currently available statistics of 23.6\,fb$^{-1}$ at the $\Upsilon$(5S)
allows to perform many interesting $B_s^0$ studies. The list 
of decays where significant signals are expected to be observed
include the following modes:

\begin{itemize}
\item $B_s^0 \to D_s^- \rho^+$ and $B_s^0 \to D_s^- a_1^+$.
\item $B_s^0 \to J/\psi \phi$, $B_s^0 \to J/\psi \eta$ and $B_s^0 \to J/\psi \eta'$.
\item $B_s^0 \to D_s^{(*)-} D_s^{(*)+}$.
\item $B_s^0 \to D_{sJ}^- \pi^+$.
\item $B_s^0 \to K^- K^+$.
\item $B_s^0 \to D^0 K^0_S$, $B_s^0 \to D^0 K^{*0}$.
\item $B_s^0 \to D_s^- \ell^+ \nu$, $B_s^0 \to D_s^{*-} \ell^+ \nu$.
\end{itemize}

Combining modes with the large number of reconstructed $B_s^0$ mesons,
the $B_s^0$ lifetime can be measured with high precision.
The production ratios for different $B^{(*)} \bar{B}^{(*)} n\pi$ and
$B_s^{(*)0} \bar{B}_s^{(*)0}$ channels can be also determined.   
It could be potentially interesting to 
take a small data sample at a higher $e^+ e^-$ CM energy,
there the $B^{**}$ or even $B_s^{**}$ signals could emerge.
A dedicated energy scan at a $\Upsilon$(5S) and $\Upsilon$(6S)
energy region is an excellent
tool to search for new $b\bar{b}$ states.
If our $\Upsilon$(5S) data sample will be increased to (100-200)\,fb$^{-1}$,
several rare $B_s^0$ decays can be potentially measured;
the most promising modes are $B_s^0 \to K^- \rho^+$, 
$B_s^0 \to \eta \eta$, $B_s^0 \to \eta \eta'$
and $B_s^0 \to \phi \phi$.

\vspace{-2.mm}
\subsection{Feasibility of {\boldmath $\Delta \Gamma_s / \Gamma_s$} measurement}

A new generation of $B$-factories with luminosity 
20-100 times higher than that delivered to the Belle and BaBar experiments
has been recently proposed \cite{sbel,sbab}. 
Taking into account an average integrated 
luminosity of about 1\,fb$^{-1}$ per day taken by Belle in first 
$\Upsilon$(5S) runs, an integrated luminosity of 1000\,fb$^{-1}$
can be taken by future $B$-factories in 10-50 days.
Although many interesting rare $B_s^0$ decays, such 
as $B_s^0 \to \gamma \gamma$ and $B_s^0 \to D_s^- K^+$,
can be measured with such a data sample, 
we would like to focus here on a few of the most important tasks. 
First of all, simple estimates indicate that the 
$\Delta \Gamma_s / \Gamma_s$ measurement can be performed with a 1\,ab$^{-1}$
data sample. 

Although the $\Delta \Gamma_s / \Gamma_s$ measurement itself is 
important, a comparison of the direct $B_s^0$ width difference
measurement with that calculated using the
${\cal B}(B_s^0 \to D_s^{(*)-} D_s^{(*)+}$) measurement is the
most important goal for such a study because it provides 
a critical Standard Model test. This method was developed 
by Grossman \cite{gross}; a more detailed description
can be found in \cite{dunie}. Technically we expect to have
$\sim$(5-7)$\%$ accuracy in
the ${\cal B}(B_s^0 \to D_s^{(*)-} D_s^{(*)+}$) measurement
with 1\,ab$^{-1}$. Therefore the accuracy of the SM test
depends mostly on the uncertainty in the
direct $\Delta \Gamma_s / \Gamma_s$ measurement.

We propose a simple method to determine the $\Delta \Gamma_s / \Gamma_s$
value based on a measurement of the distance between two $B_s^0$ vertices.
We assume fully anti-correlated $CP$ values of the two final state 
$B_s^0$ mesons in the decay channel 
\mbox{$\Upsilon$(5S) $\to B_s^* \bar{B}_s^*$}.
To estimate the uncertainty of $\Delta \Gamma_s / \Gamma_s$ measurement
with 1\,ab$^{-1}$ of data at the $\Upsilon$(5S), we use a toy MC simulation
to model the distance between two $B_s^0$ decay vertices (Fig.~10).
The number of fully reconstructed $B_s^0$ mesons with a data sample
of 1\,ab$^{-1}$ is estimated to be effectively $\sim$4000 events.
This sample is expected 
to comprise $\sim$1000 $D_s^{(*)-} D_s^{(*)+}$ events,
$\sim$1500 $J/\psi\ \eta/\eta'$ events, $\sim$600 $K^+K^-$ events
and $\sim$2200 $J/\psi \phi$ events. The $J/\psi \phi$ events
contain $\sim 20\%$ $CP$-odd and $\sim 80\%$ $CP$-even components, and
we reduce this effective event number to $\sim$1000.

Assuming $CP$ conservation in $B_s^0$ decays, the mass eigenstates 
are $CP$-eigenstates,
and one could compare the lifetime distributions of a $B_s^0 /\bar{B}_s^0$
decaying to $CP$-even and $CP$-odd final state.
To measure the $\Delta \Gamma_s = \Gamma_L - \Gamma_S$ value experimentally
($\Gamma_L$ and $\Gamma_S$ are widths of long-lived 
and short-lived $B_s^0$ mesons),
we should select all reconstructed $B_s^0$ decays with fixed $CP$, 
and determine the decay vertices for this $B_s^0$ with fixed $CP$ 
and for the second $B_s^0$ in the event.
Then we can measure the distance between the decay vertices
$\Delta Z = Z_{CP even} - Z_{CP odd}$ along the $\Upsilon$(5S) boost 
direction and, respectively, the time difference 
$\Delta t = \Delta Z / (\beta \gamma c)$
(here $\beta \gamma$ is the boost factor).
The expected distribution is proportional to 
$exp(-\Gamma_L \Delta t)$ for $\Delta t > 0$ and 
$exp(+\Gamma_S \Delta t)$ for $\Delta t < 0$; thus, fitting to this
distribution yields both $\Gamma_L$ and $\Gamma_S$.

\begin{figure*}[t]
\centering
\includegraphics[width=50mm]{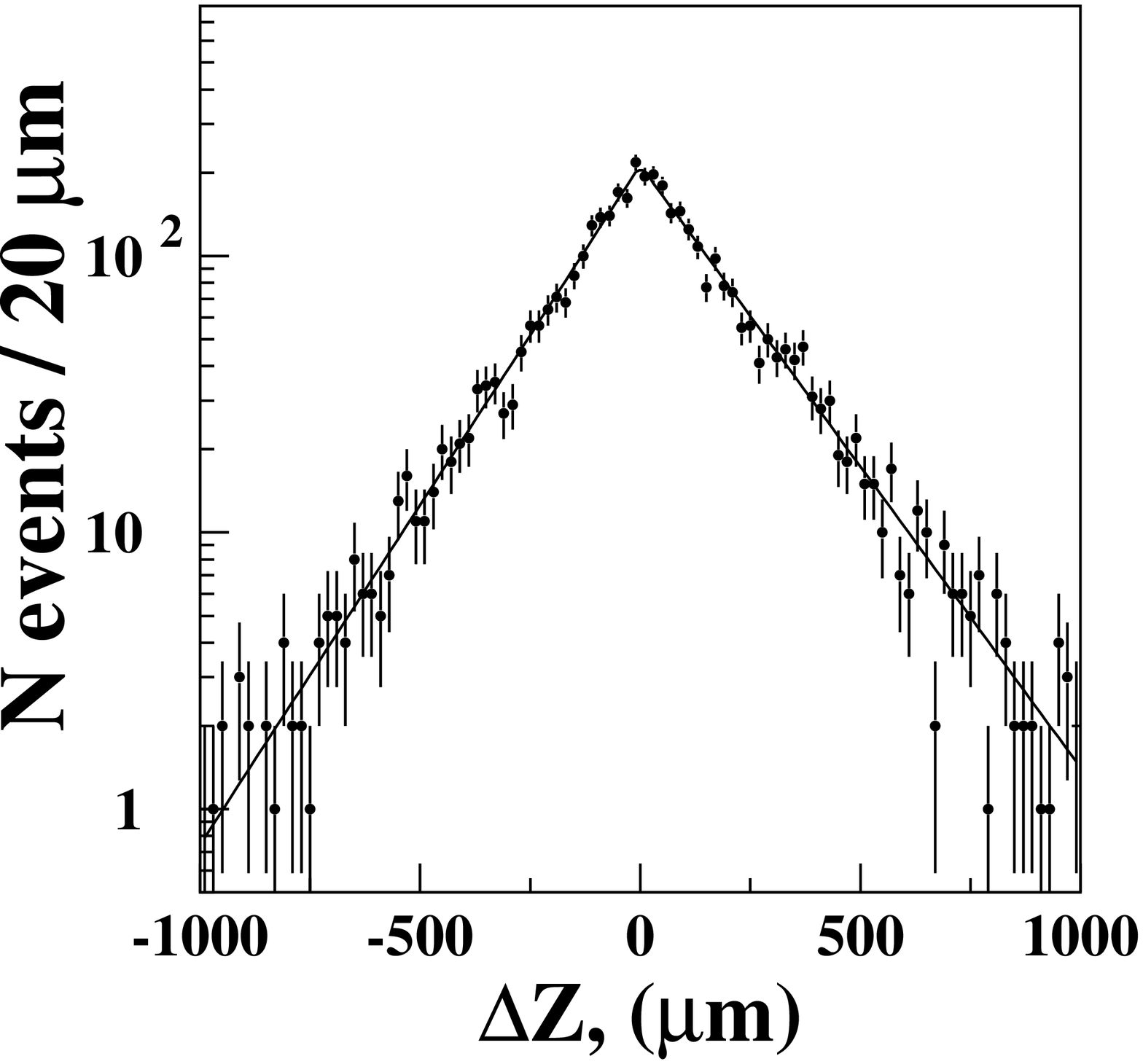}\hspace{0.5cm}\includegraphics[width=50mm]{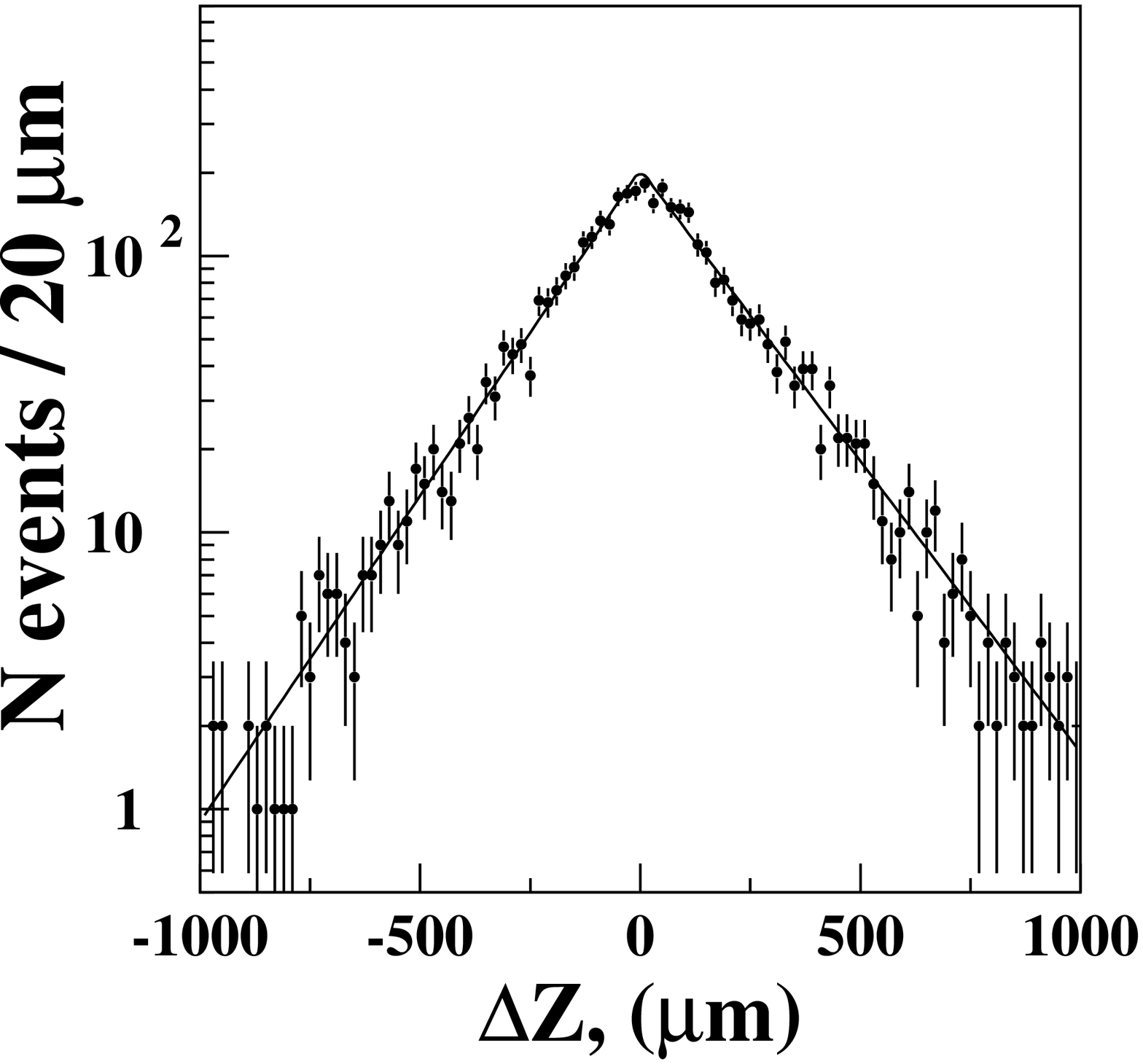}
\vspace{-0.1cm}
\caption{The MC simulated $\Delta Z$ distributions 
between two $B_s^0$ vertices are shown for the single vertex resolution
neglected (left) or modeled by a Gaussian with $\sigma=$ 40$\mu m$ (right).
Full curves show the results of fits described in the text.}
\end{figure*}

Finally, from the $\Delta Z$ distribution fit we obtained the 
value $\Delta \Gamma_s / \Gamma_s = (10.2 \pm 2.4)\%$ 
(significance \mbox{$\sim\,4\sigma$}), where value
$\Delta \Gamma_s / \Gamma_s = 10.0 \%$ was fixed in the toy MC simulation.
Because background contribution is low in the studied modes,
the signal significance depends on statistics as $\propto\sqrt{L_{\rm int}}$.
The vertex resolution effect is negligible: almost no degradation
in accuracy of $\Delta \Gamma_s$ measurement was found
comparing the $\Delta Z$ distribution fit values with the vertex resolution
neglected (Fig.~10, left) in toy MC simulation and the single vertex 
resolution of $\sim40\mu m$ (Fig.~10, right) included in toy MC, 
latter reproduces roughly
the current Belle vertex reconstruction uncertainty in $\Delta Z$.
It is interesting that we can use the event numbers $N^+$
and $N^-$ in the $\Delta Z > 0$ and $\Delta Z < 0$ regions, respectively, 
to estimate
$\Delta \Gamma_s / \Gamma_s = 2 \times ( N^+ - N^- )\ /\ ( N^+ + N^- )$,
however the statistical uncertainty is $\sim 30\%$ larger in this method.

\subsection{Feasibility of {\boldmath $B_s^0-\bar{B}_s^0$} mixing measurement
with improved vertex resolution}

Theoretically almost zero $CP$ violation is expected in $B_s^0$ mixing
within the Standard Model, therefore, the search for the $CP$ violation
provides an important opportunity to observe effects Beyond the Standard Model.
It is interesting to note that the conventional method of time dependent
$CP$ violation measurement is often assumed to be impossible 
at the $\Upsilon$(5S) due to very fast $B_s^0$ oscillations.
However this may be possible, since the corresponding distance 
between oscillation function maximum and minimum at the $\Upsilon$(5S) is 
\mbox{$D = \pi \cdot \Delta m_s \cdot \beta \gamma c = 22.5\,\mu$m}.
Such decay vertex resolution can be reached by existing vertex detectors:
currently discussed Super Belle single vertex resolution (SVR)
is expected to be $\sim 20\,\mu$m for tracks with momentum greater
than 1.5\,GeV/c and direction nearly perpendicular to the beam axes.

The $B_s^0$ oscillations can be observed using the same-sign (SS) and
opposite-sign (OS) high momentum leptons (their intersections 
with beam profile can be used to obtain vertex positions) 
produced in two semileptonic $B_s^0$ decays from 
the $\Upsilon {\rm (5S)} \to B_s^{(*)0} \bar{B}_s^{(*)0}$ decay.
The number of SS and OS events is very large in 1\,ab$^{-1}$ data sample
and is not critical for such an analysis.
In Fig.~11 the MC simulated $\Delta Z$ distributions
with 1\,ab$^{-1}$ dataset are shown for (from left to right)
SS lepton events with SVR = $10\,\mu$m, SS-OS event difference
with SVR = $10\,\mu$m, SS lepton events with SVR = $13\,\mu$m, and
SS-OS event difference with SVR = $13\,\mu$m.
Oscillations are perfectly seen with SVR = $10\,\mu$m,
but the sinusoidal shape begins to disappear at SVR = $13\,\mu$m.
Therefore, to measure $B_s^0$ mixing at Super Belle 
the SVR should be improved to $\sim15\,\mu$m or 
a $\Upsilon$(5S) boost value should be increased by at least 25$\%$.

\begin{figure*}[h]
\centering
\includegraphics[width=40mm]{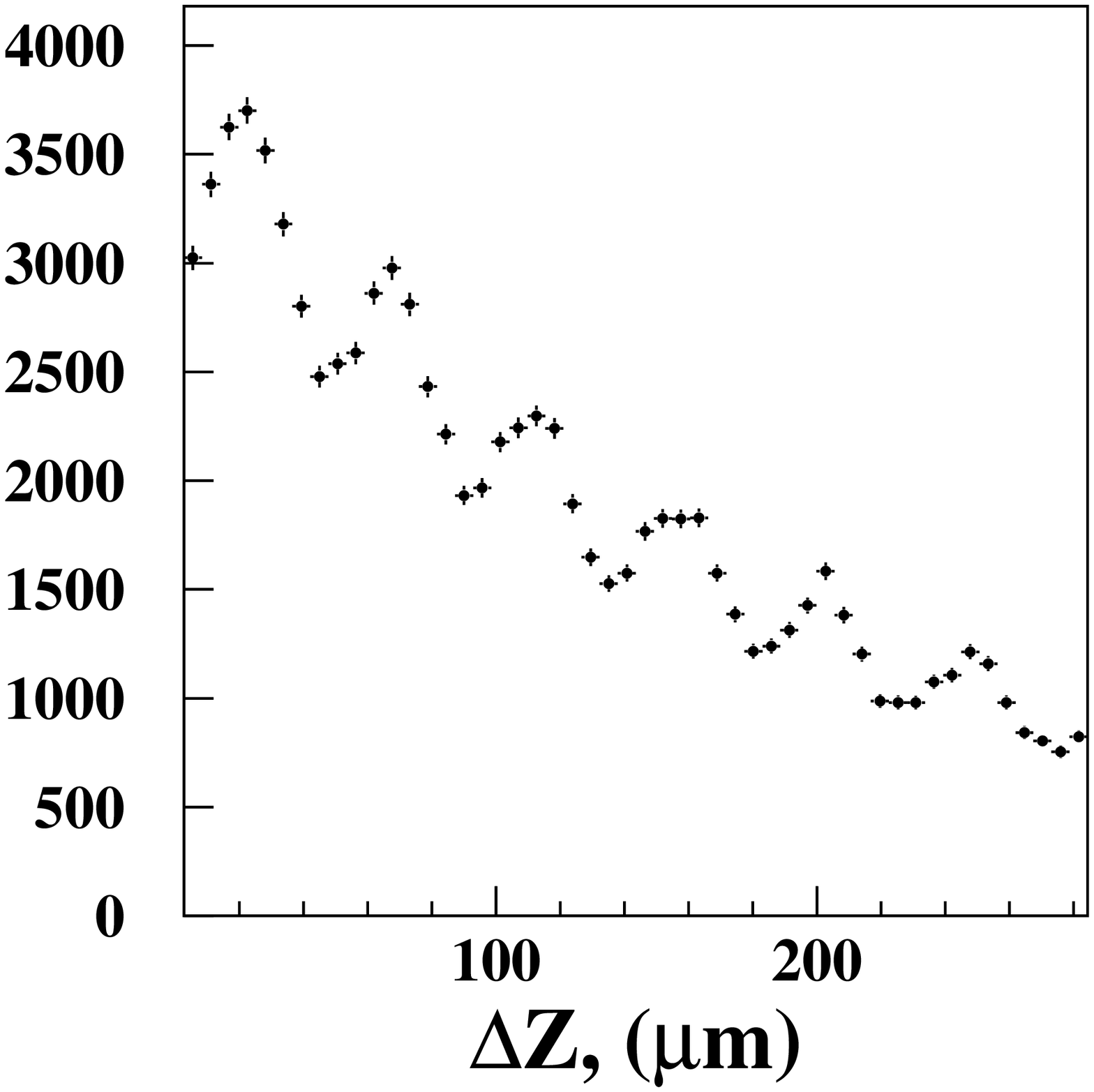}\hspace{0.1cm}\includegraphics[width=40mm]{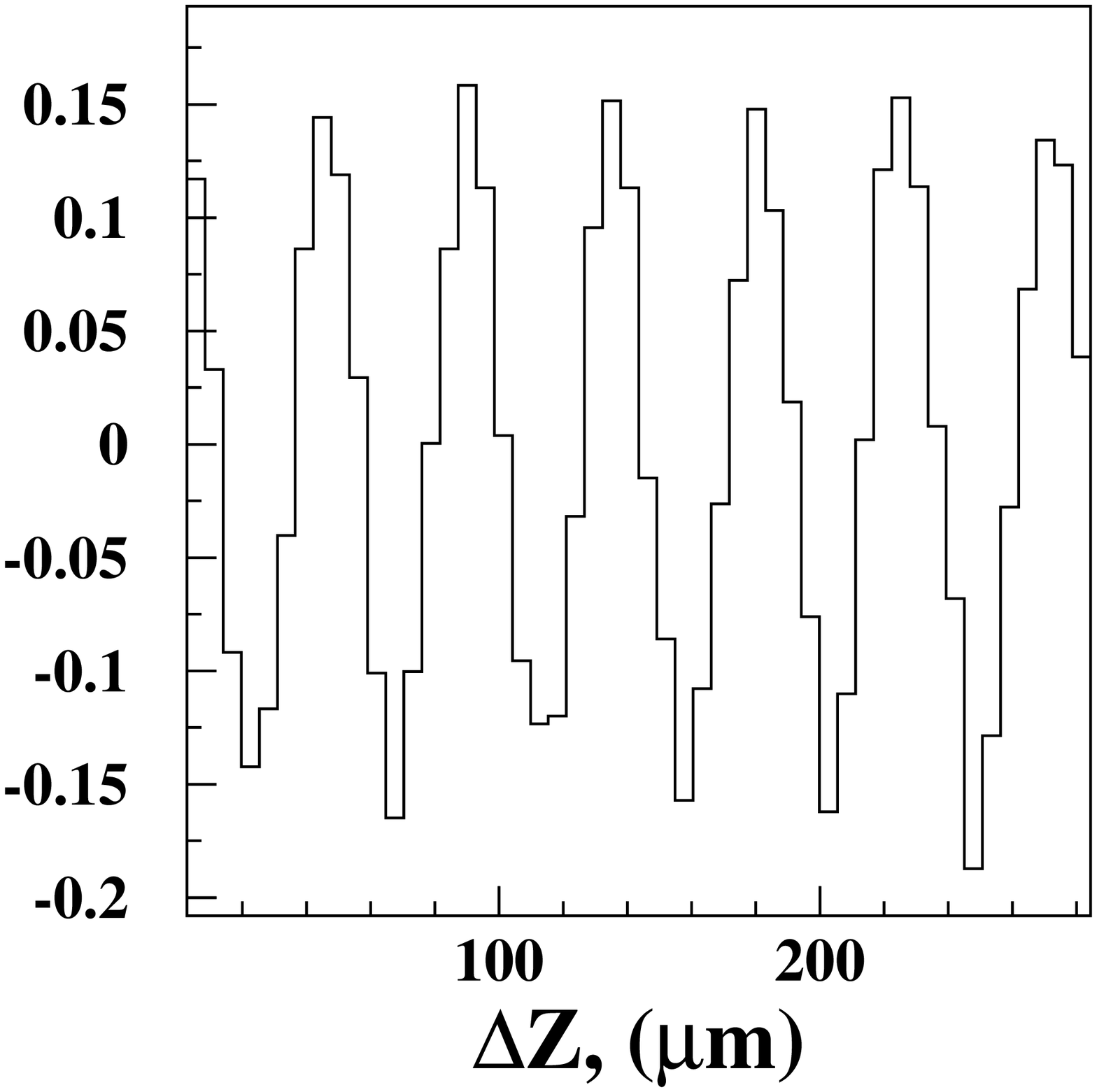}\includegraphics[width=40mm]{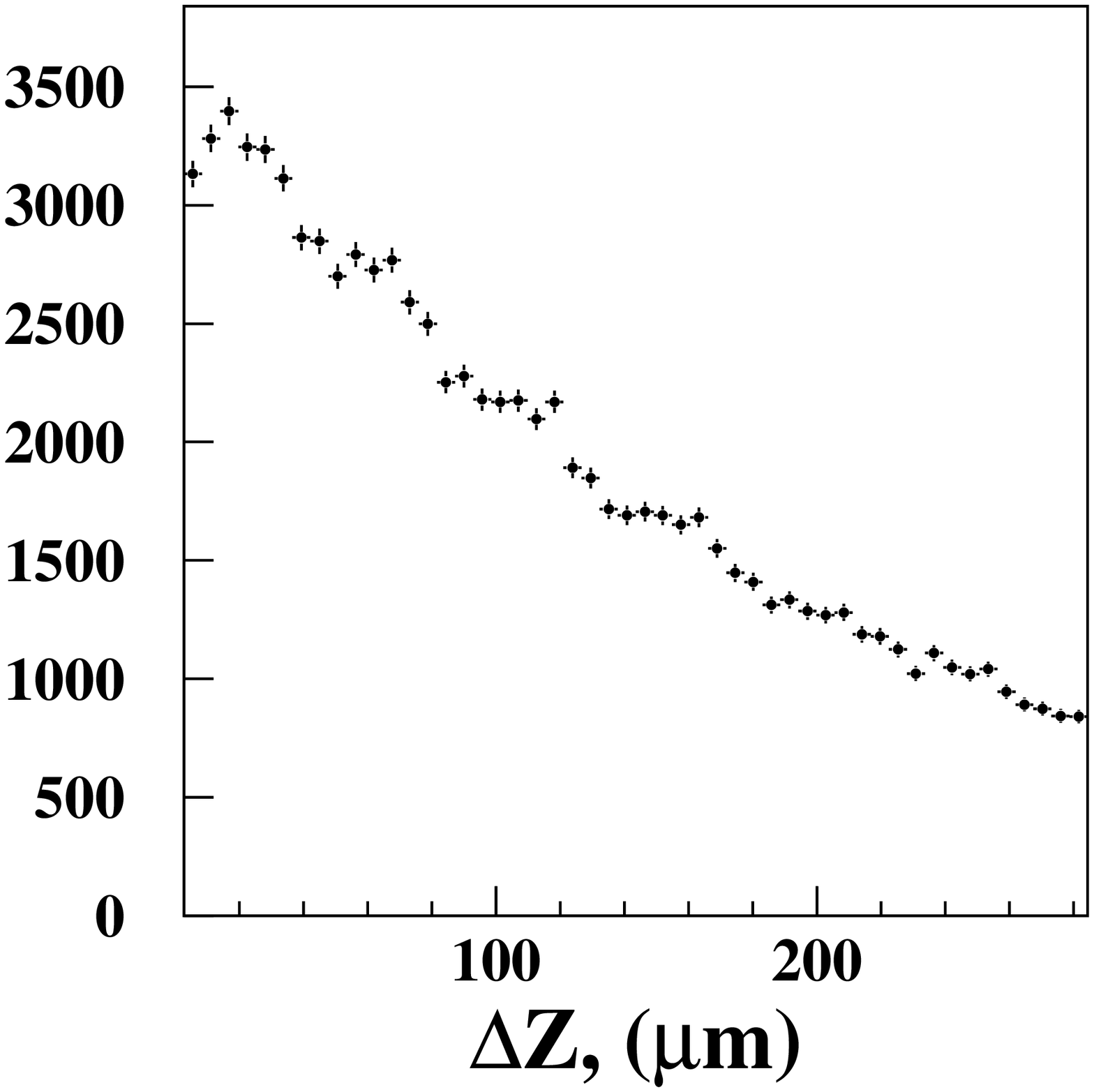}\hspace{0.1cm}\includegraphics[width=40mm]{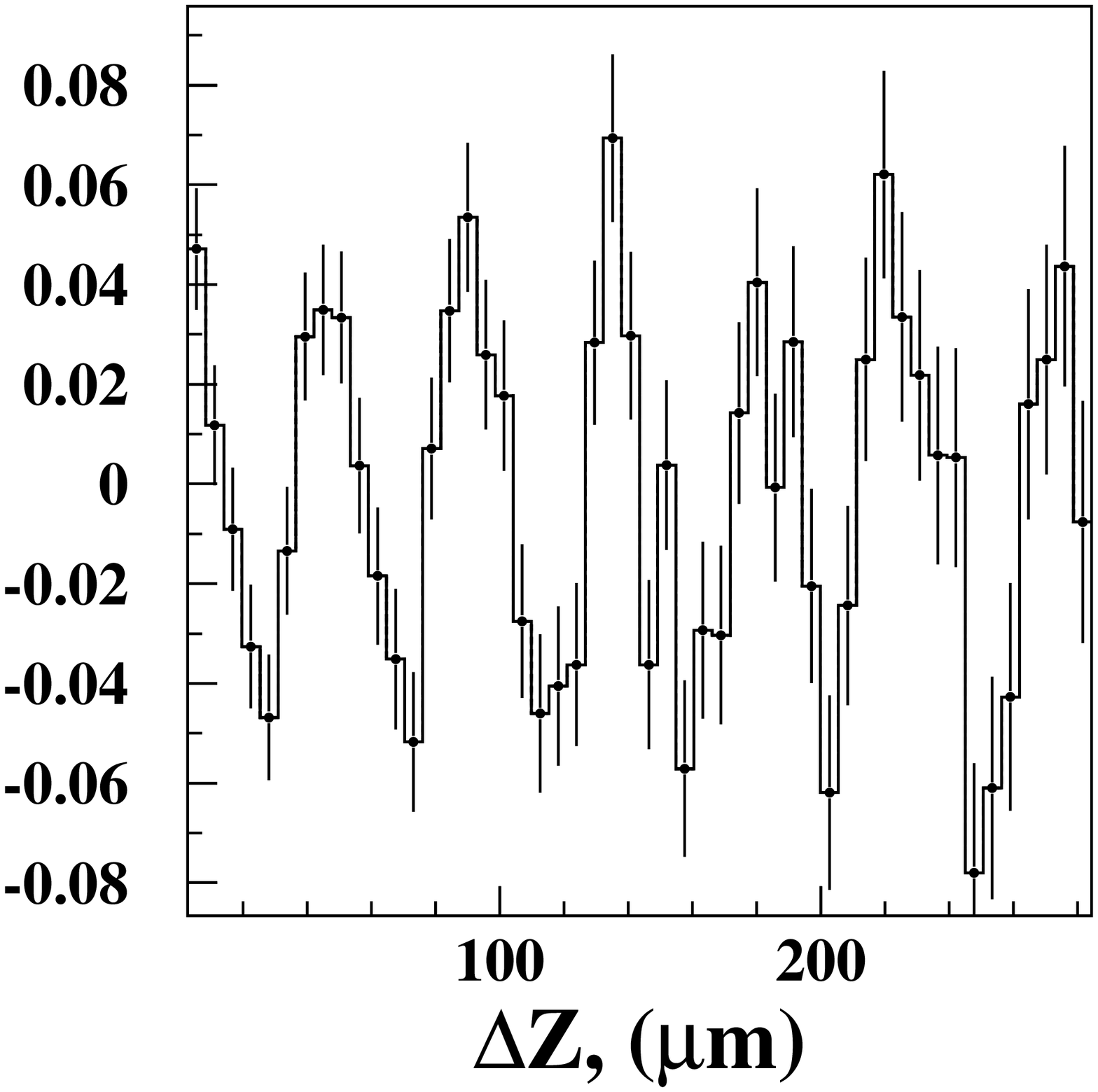}
\vspace{-0.1cm}
\caption{The event numbers for SS leptons (first) and event number 
difference for SS and OS leptons (second) as a function of
$\Delta Z$ is shown
for SVR = $10\,\mu$m (two left), and for SVR = $13\,\mu$m (two right).
The distributions are simulated using toy MC.}
\end{figure*}

\subsection{Potential measurements with higher beam energies}

One of the significant advantages of hadron-hadron colliders is
the possibility to study many heavy beauty states, such as
$B^{**}$, $B_s^{**}$, $B_c$, $\Lambda_b$, $\Sigma_b$ et. al.
Potentially, however, all these states can be studied 
at the $e^+ e^-$ collider running at the center-of-mass energy
(12-14)\,GeV. Even outside $\Upsilon$ resonances, the
$b\bar{b}$ continuum is expected to be $\sim 10\%$ of all
continuum events. In particular, the $\Lambda_b$ 
baryons could be 
produced at the center-of-mass energy of $\sim 11.3\,$GeV
(threshold for $\Lambda_b \bar{\Lambda}_b$ pair production 
is $E=11.248\,$GeV) with approximately the same rate as $B_s^0$ mesons.
In a very high luminosity $B$-factory with a wide allowed
beam energy range, the study of $B_c$ mesons
seems to be also possible.

In conclusion, we discussed recent results and future prospects 
for the $\Upsilon$(5S) running at $B$-factories. 
Many new results are obtained with the 23.6\,fb$^{-1}$
data sample collected by Belle at the $\Upsilon$(5S)
and more results are expected soon. The results obviously
demonstrate that $B_s^0$ meson studies at the $\Upsilon$(5S) have a
great potential. 
It is important to have a large allowed range of beam energies
and a high performance vertex detector at future Super $B$-factories.
Important tests of the Standard Model can be 
performed with statistics of the order of 1\,ab$^{-1}$ at the $\Upsilon$(5S).

\end{document}